\documentclass[fleqn,usenatbib]{mnras}

\usepackage[T1]{fontenc}

\DeclareRobustCommand{\VAN}[3]{#2}
\let\VANthebibliography\thebibliography
\def\thebibliography{\DeclareRobustCommand{\VAN}[3]{##3}\VANthebibliography}

\usepackage{graphicx}	
\usepackage{amsmath}	
\usepackage{amssymb}	
\usepackage{siunitx}
\usepackage{multirow}
\usepackage{newtxtext,newtxmath}

\newcommand{\um}[1]{\SI{#1}{\micro\meter}}

\title[TNG50 galaxy luminosity functions]{UV to submillimetre luminosity functions of TNG50 galaxies}

\author[A. Tr\v{c}ka et al.]{Ana Tr\v{c}ka,$^{1}$\thanks{E-mail: anatrcka@hotmail.com}
Maarten Baes,$^{1}$
Peter Camps,$^{1}$
Anand Utsav Kapoor,$^{1}$ 
Dylan Nelson,$^{2}$
Annalisa Pillepich,$^{3}$  \newauthor
Daniela Barrientos,$^{1}$
Lars Hernquist,$^{4}$
Federico Marinacci,$^{5}$ and
Mark Vogelsberger$^{6}$
\\
$^{1}$Sterrenkundig Observatorium, Universiteit Gent, Krijgslaan 281 S9, B-9000 Gent, Belgium\\
$^{2}$Universit\"at Heidelberg, Zentrum f\"ur Astronomie, Institut f\"ur theoretische Astrophysik, Albert-Ueberle-Str. 2, 69120 Heidelberg, Germany\\
$^{3}$Max-Planck-Institut f\"ur Astronomie, K\" onigstuhl 17, 69117 Heidelberg, Germany\\
$^{4}$Center for Astrophysics | Harvard \& Smithsonian, 60 Garden Street, Cambridge, MA 02138, USA\\
$^{5}$Department of Physics \& Astronomy 'Augusto Righi', University of Bologna, via Gobetti 93/2, I-40129 Bologna, Italy\\
$^{6}$Department of Physics, Kavli Institute for Astrophysics and Space Research, Massachusetts Institute of Technology, Cambridge, MA 02139, USA
}

\date{Accepted XXX. Received YYY; in original form ZZZ}

\pubyear{2022}

\begin{document}
\label{firstpage}
\pagerange{\pageref{firstpage}--\pageref{lastpage}}
\maketitle

\begin{abstract}
We apply the radiative transfer (RT) code SKIRT on a sample of $ \sim 14 000$ low-redshift ($z\le 0.1$) galaxies extracted from the TNG50 simulation to enable an apples-to-apples comparison with observations. The RT procedure is calibrated via comparison of a subsample of TNG50 galaxies with the DustPedia observational sample: we compare several luminosity and colour scaling relations and spectral energy distributions in different specific SFR bins. 
We consistently derive galaxy luminosity functions for the TNG50 simulation in 14 broadband filters from UV to submillimetre wavelengths and investigate the effects of the aperture, orientation, radiative transfer recipe, and numerical resolution.
We find that, while our TNG50+RT fiducial model agrees well with the observed luminosity functions at the knee ($\pm$ 0.04 dex typical agreement), the TNG50+RT luminosity functions evaluated within $5\,R_{1/2}$ are generally higher than observed at both the faint and bright ends, by 0.004 (total IR)-0.27 (UKIDSS H) dex and 0.12 (SPIRE250)-0.8 (GALEX FUV) dex, respectively. 
A change in the aperture does affect the bright end of the luminosity function, easily by up to 1 dex depending on the choice. 
However, we also find that the galaxy luminosity functions of a worse-resolution run of TNG50 (TNG50-2, with 8 times worse mass resolution than TNG50, similar to TNG100) are in better quantitative agreement with observational constraints.
Finally, we publicly release the photometry for the TNG50 sample in 53 broadbands from FUV to submillimetre, in three orientations and four apertures, as well as galaxy spectral energy distributions.
\end{abstract}

\begin{keywords}
methods: numerical -- submillimetre: galaxies -- galaxies: evolution -- galaxies: formation -- ISM: dust, extinction -- radiative transfer
\end{keywords}



\section{Introduction}

The Universe is an immensely sophisticated and exceedingly complex system, and understanding the structure, formation and evolution of its ingredients is a daunting job.
In recent years, there has been a considerable improvement in the tools that can aid in tackling this difficult task, and one of these are cosmological simulations \citep[for a review, see][]{Vogelsberger2020a}.
Cosmological galaxy hydrodynamical simulations, which incorporate baryons as well as dark matter, come in two flavors: `zoom-in' and `large-volume'.
While zoom-in simulations focus on individual galaxies and can explore small-scale processes like star formation, interstellar medium physics, stellar and active galactic nuclei (AGN) feedback in greater depth, they usually lack the required numbers for a statistical analysis \citep{Guedes2011, Sawala2016, Wetzel2016, Grand2017, Tremmel2019, Font2020}.
On the other side, the large-volume simulations simultaneously produce various galaxy populations in abundance, at the cost of details on galaxy inner processes \citep{Dubois2014, Vogelsberger2014b, Schaye2015, Pillepich2018a, Dave2019}. 

To assess the realism of the simulations, and thereby also constrain the values of their various model parameters, in order to use them to describe and predict physical phenomena, simulations have to be compared with observations.
This poses a challenge, as it is necessary to convert one into the domain of the other, e.g. light to mass with observed data (`inverse modelling') or vice versa through the creation of synthetic observations from simulated data (`forward modelling').
For the purposes of observations of galaxies, inverse modelling relies on a set of (typically poorly constrained) assumptions including mass-to-light ratio, star formation history and dust attenuation curve, which can introduce biases in the derived physical properties \citep{Mitchell2013,LoFaro2017}. 
The alternative solution, the forward modelling approach, can alleviate these, while incorporating the complex galaxy geometry at the same time.

Galaxy luminosity functions (LF), defined as a galaxy count per unit volume and unit luminosity, are a fundamental property of a galaxy population. 
In particular, LFs at low redshift are easy to access observationally and they provide a benchmark and impose constraints on the buildup of galaxies at earlier times. LFs at different wavelengths offer distinct information on the galaxy properties. The UV LF \citep[e.g.,][]{Sullivan2000, Budavari2005, Wyder2005} describes the distribution of unobscured star formation. Fluxes in the optical and especially the near-infrared (NIR) trace the evolved stellar populations, making these LFs \citep[e.g.,][]{Bell2003, Blanton2003, Hill2010, Loveday2012, Driver2012} proxies for the galaxy stellar mass function, which is directly available from the simulations and is typically used for their calibration or design \citep[e.g.][]{Dubois2014, Vogelsberger2014b, Schaye2015, Pillepich2018a, Dave2019}.
Thermal emission by dust, often used as a tracer for obscured star formation, can be probed using infrared (IR) LFs \citep[e.g.][]{Vaccari2010, Dunne2011, Negrello2013, Marchetti2016}.
Analysing the LFs simultaneously in different bands can provide detailed insights into galaxy formation and evolution, and comparing LFs from simulated and observed data can evaluate the current and instruct and constrain future galaxy formation modeling phase.

To derive galaxy luminosity functions from cosmological hydrodynamical simulations, the forward modelling approach is necessary. An important ingredient in any forward modelling method is interstellar dust. It typically absorbs $30\%$ of the stellar light in star-forming galaxies and converts it to thermal infrared radiation \citep{Viaene2016,Bianchi2018}. However, while advancing, cosmological hydrodynamical simulations generally still do not model dust on-the-fly as a separate component from gas \citep[but see e.g.][]{McKinnon2018,Dave2019,Granato2021}. So, in order to incorporate this important component in a forward modelling approach, the total dust mass, physical properties and spatial distribution have to be further modeled in post-processing based on local gas properties and/or global galaxy properties.
One option is to perform the modelling by assuming a simple dust geometry \citep[e.g., a screen, slab or a simple dusty disc][]{Calzetti2000,Charlot2000,Yuan2021}, to account for the effect of dust attenuation \citep{Trayford2015,Trayford2020, Dickey2021, Hahn2021,Tang2021}. 

A more advanced option is to apply a full 3D dust radiative transfer (RT) procedure \citep[e.g.,][]{Camps2016, Camps2018, Schulz2020, Shen2021, Kapoor2021, Lovell2021,Popping2022}. This has the advantage that the intrinsic 3D star-dust geometry can be incorporated, which increases the physical fidelity of the results \citep{Trayford2017, Nelson2018, Rodriguez-Gomez2019}. In typical RT recipes, the compact dust is traced by the young stellar particles, while the diffuse dust is derived from the gas properties in the simulation \citep[e.g.][]{Jonsson2010, Camps2016, Camps2018, Trayford2017, Vogelsberger2020b,Schulz2020, Millard2021}.

Galaxy luminosity functions at low-redshift have been compared between a number of simulations and observational samples. Examining the properties of the Horizon-AGN simulations  \citep{Dubois2014}, \citet{Kaviraj2017} derived $K$ and $r$-band LFs at $z=0.1$ using a screen dust model and found that the simulation results in the $K$ band overshoot the observations, while underpredicting them in the $r$ band.
\citet{Trayford2015} calculated UV to NIR LFs for a galaxy sample extracted from the EAGLE simulations \citep{Schaye2015, Crain2015} at $z=0.1$, applying a two-screen model to account for the dust effects.   
They found overall agreement with the observations, with a slight galaxy excess at the faint end and an underestimation at the bright end.
\citet{Baes2020} derived the IR LFs at $z<0.2$ by combining two EAGLE simulation volumes and therefore resolutions \citep{Schaye2015}, to properly sample both ends of the LF. The mock IR data were obtained using a RT procedure \citep{Camps2016, Camps2018, Trayford2017}, calibrated on the Herschel Reference Survey \citep[HRS;][]{Boselli2010, Cortese2012}.  
They found reasonable agreement with the observed LFs, with an underestimation of the high IR bright galaxies.

In this paper, we exploit the state-of-the-art cosmological hydrodynamical simulation TNG50 \citep{Nelson2019b,Pillepich2019}, the highest-resolution version of the suite of IllustrisTNG simulations. 
We perform a RT post-processing procedure on a sample of $\sim 14 000$ low-redshift ($z\le 0.1$) galaxies with $M_{\mathrm{star}}>10^8 \mathrm{M_\odot}$. 
The procedure is calibrated by comparing a subset of TNG50 galaxies with the DustPedia sample \citep{Davies2017} of nearby galaxies in terms of several scaling relations. The calibration is carried out by applying the spectral energy distribution (SED) fitting tool \textsc{cigale} \citep{Boquien2019} on both observed and simulated fluxes, to guarantee that the comparison is self-consistent \citep{Trcka2020}.
Next, we uniformly construct galaxy luminosity functions for the low-redshift simulated sample in 14 broadband filters, from the UV to the submillimetre (submm) range. Finally, we provide fluxes and absolute magnitudes for the whole sample and for a number of apertures and galaxy orientations, which will be useful for a diversity of future studies.

We organise the paper as follows. In Sec. \ref{sec:method}, we
define the \textsc{skirt} RT post-processing procedure and present the calibration of the RT procedure. In Sec. \ref{subsec:LF} we show LFs of the TNG50 simulations at a range of wavelengths from UV to submm.
Our results are discussed in Sec. \ref{sec:disc} and summarised in Sec. \ref{sec:conc}.

\section{Method}
\label{sec:method}

\subsection{TNG50 simulation}
\label{sec:TNG} 

IllustrisTNG \citep[][hereafter TNG]{Pillepich2018b, Pillepich2019,Nelson2018,Nelson2019b,Naiman2018,Marinacci2018,Springel2018}, the successor of the Illustris simulation \citep{Vogelsberger2014a,Vogelsberger2014b,Genel2014}, represents the current state of the art in cosmological hydrodynamical simulations. 
While both Illustris and TNG rely on the moving-mesh code  \textsc{arepo} \citep{Springel2010} as the underlying hydrodynamical solver, the galaxy formation prescriptions for TNG have been upgraded, including modifications of the galactic winds, AGN feedback, stellar evolution and the incorporation of magnetic fields \citep{Weinberger2017,Pillepich2018a}.
The TNG simulation data is publicly available \citep{Nelson2019b}.

From a range of available volumes and resolutions, for this study, we opted for the high-resolution TNG50 simulation with a medium volume which evolves a cubic volume of 50 Mpc on a side \citep{Nelson2019a, Pillepich2019}. 
This simulation offers the best trade-off between resolution and volume, as the baryonic mass resolution of $8.5\times  10^4 \mathrm{M_\odot}$ and the typical star-forming gas cell of $\sim 140$ pc correspond to typical zoom-in simulations, while the simulation volume ensures a considerable number of differently sized galaxies for a proper statistical analysis.
The cosmological parameters are based on the \citet{Planck2016} results ($\Omega_m = 0.3089$, $\Omega_b = 0.0486$, $\Omega_\Lambda = 0.6911$, $H_0 = \mathrm{100 \; h \; km \; s^{-1}Mpc^{-1}}$ with $h = 0.6774$) and the initial mass function (IMF) is from \citet{Chabrier2003}.
We adopt the same parameters throughout this study, unless stated otherwise.

\begin{figure}
 	\includegraphics[width=\columnwidth]{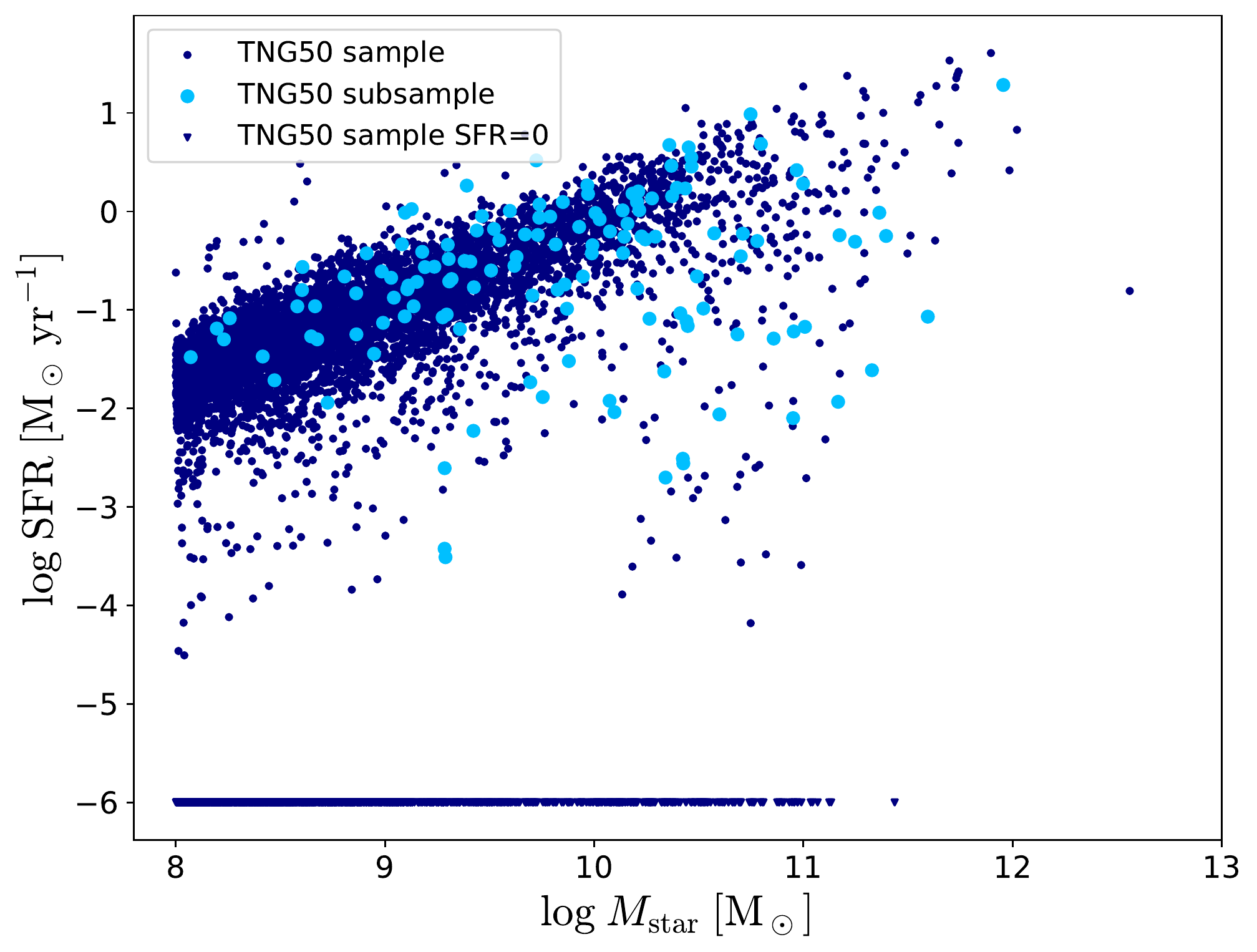}
    \caption{The main sequence of the $z=0$ TNG50 sample used in this study (navy blue). The triangles with SFR of $10^{-6}\mathrm{M_\odot/yr}$ are those with unresolvable low levels of SFR. The TNG50 subsample used for the RT calibration against the DustPedia observed galaxies is shown in cyan.}
    \label{fig:select}
\end{figure}

To ensure a sufficient number of stellar particles per galaxy ($\gtrsim 10^3$), we limit our analysis to galaxies with a stellar mass inside twice the stellar half mass radius ($R_{1/2}$ hereafter) of at least $10^8 M_\odot$.
We focus our study on low redshifts, so this leaves us with 7375 galaxies at $z=0$ and 7302 at $z=0.1$. 
The location of these galaxies at $z=0$ in the $M_\star-\mathrm{SFR}$ plane is shown in Fig.~\ref{fig:select} in navy blue. 
The triangles represent 1588 galaxies with unresolvable low levels of SFR i.e. $\mathrm{SFR} < 10^{-6} \mathrm{M_\odot/yr}$ \citep{Donnari2019}.

\subsection{The RT procedure}
\label{sec:RT}

The RT post-processing of the TNG50 galaxies is based on the 3D Monte Carlo RT code \textsc{skirt} \citep{Baes2011, Camps2015, Camps2020} (Version 9). 
The code treats the relevant processes of dust extinction, emission, scattering and self-absorption \citep{Camps2015}. \textsc{skirt} is fully parallelised \citep{Verstocken2017} and it uses a suite of optimisation techniques to enhance efficiency \citep{Baes2011,Baes2016}. It is equipped with an extensive library of SED models, dust models and geometry models \citep{Baes2015} and advanced spatial grids \citep{Saftly2013, Saftly2014, Camps2013}. 
In the past few years, \textsc{skirt} has been extensively used to generate synthetic observations for hydrodynamically simulated galaxies, including synthetic fluxes and SEDs 
\citep[e.g.,][]{Camps2016, Trayford2017, Camps2018, Vogelsberger2020b}, images
\citep[e.g.,][]{Saftly2015, Rodriguez-Gomez2019, Camps2022, Popping2022}, and polarisation maps
\citep{Vandenbroucke2021}.

\subsubsection{Galaxy partitioning}
\label{subsec:gal_part}

The strategy applied here largely follows the previous work on the EAGLE \citep{Camps2016, Camps2018, Trayford2017} and Auriga \citep{Kapoor2021} simulations, albeit with a number of minor adjustments. For every TNG50 galaxy, we analyse three sources of light: evolved stars, the star-forming (SF) regions, and the diffuse dust. 

\paragraph*{Evolved stellar populations:}
Information on the star particles is obtained directly from the simulation. To each star particle that is older than 10 Myr, we assign an SED from the \cite{Bruzual2003} family, based on the particle birth mass, metallicity, and age. The \cite{Chabrier2003} IMF is assumed. For the smoothing length needed for the RT, we use the radius of the sphere surrounding 32 stellar particles centred on each particle. 
Test results corresponding to different values for the smoothing length are described in Appendix~\ref{sec:A-tests}. We argue that the adaptive smoothing length we adopt gives the most realistic results \citep{Rodriguez-Gomez2019, Schulz2020}, however, there is no single optimal value.

\paragraph*{Star-forming regions:}
The modelling of the star-forming regions includes the use of the \textsc{mappings-iii} SED template family \citep{Groves2008}, applied on the young star particles ($t$<10 Myr).
The templates are typically used for this purpose \citep[e.g.][]{Camps2015,Trayford2015,Camps2018,Schulz2020,Kapoor2021} and they contain five free parameters: interstellar medium (ISM) pressure, compactness, SFR,  photodissociation region (PDR) covering fraction, and metallicity. 
\begin{itemize}
    \item For the ISM pressure, we assume the value of $P/k=10^5 \; \mathrm{K~cm^{-3}}$, where $k$ is the Boltzmann constant. This is an average value in the \textsc{mappings-iii} library \citep{Groves2008}, and a representative value given the relevant densities in the ISM in the TNG50 simulation galaxies and the effective ISM temperature \citep{Springel2003}. 
    The specific choice of the ISM pressure value hardly affects the broadband fluxes \citep[see][their Fig.~4]{Groves2008}.
    \item The compactness parameter ($\cal C$) mostly regulates the FIR part of the spectrum, with higher values corresponding to the shift of the IR peak towards shorter wavelengths, or equivalently, towards higher dust temperatures. Similarly as we did for the pressure, we assume the typical value $\log{\cal C}=5$. 
    However, we do not use the same value for all SF regions, but we sample from the Gaussian distribution with the mean 5 and $\sigma = 0.4$ \citep{Kapoor2021}. 
    This results in a more realistic blend of SEDs with different dust temperatures, similar to observed or simulated star forming regions \citep{Utomo2019,Kannan2020}.
    \item The SFR is calculated from the birth mass of the particle over 10 Myr, since the SED templates are assuming the constant SFR over 10 Myr. 
    \item The PDR covering fraction ($f_{\mathrm{PDR}}$) describes to what extent the H\textsc{ii} regions are coated with the PDRs. 
    As we possess the information on the particle age ($t$), and since it is expected that the PDRs will clear eventually, we adopt a parameter $\tau$ such that 
    \begin{align}
    f_{\mathrm{PDR}}=e^{-t/\tau}
        \label{eq:tau}
    \end{align}
    \citep[and references therein]{Groves2008}. 
    If $\tau=0$ the SF region is completely transparent, while if $\tau=\infty$, the H{\sc{ii}} region is completely covered with the PDR. Since the particle age is a discrete value, we perturb it slightly\footnote{More precisely, we sample from the Gaussian with the particle age as mean and $\sigma=0.2\; \mathrm{Myr}$.} to introduce some diversity for the SF regions with the same age. 
    This can affect the median galaxy $f_{\mathrm{PDR}}$ up to 0.05.
    We use $\tau$ as a free parameter. 
    \item The metallicity is obtained from the simulation.
\end{itemize}

The \textsc{mappings-iii} templates use Starburst99 family of SEDs for single stellar populations (SSP) \citep{Leitherer1999} and the \citet{Kroupa2001} IMF, while for the older stellar populations, we use the \cite{Bruzual2003} SSP templates and the \cite{Chabrier2003} IMF. 
While we could use this SSP library (and therefore also the \citet{Kroupa2001} IMF) for the evolved stars as well, we opted not to, as this would produce a significant inconsistency with the IMF from the simulations, and later with the SED fitting.
To investigate the actual difference between the two SSP libraries, we ran additional \textsc{skirt} simulations where we adopt only the Starburst99 templates for all stellar particles (young and evolved). 
The results of the test are shown in Fig.~\ref{fig:all_tests}d. 
The median magnitude deviation for the 136 TNG50 galaxies has a maximum of 0.23 mag. This in turn would affect optical luminosities up to 0.1 dex.

\paragraph*{Diffuse dust}
\label{subsubsec:diff_dust}
The diffuse dust component is derived from the gas simulated by TNG50. To select gas cells eligible to contain dust, we apply the boundary from \citet{Torrey2012, Torrey2019} that separates the hot circumgalactic medium from the cooler gas, 
\begin{equation}
\log \left(\frac{T_{\text{gas}}}{\text{K}}\right) 
= 
6 + 0.25\log \left(\frac{\rho_{\text{gas}}}{10^{10}~h^2~{\text{kpc}}^{-3}}\right),
\label{T12}
\end{equation}
where $T_{\text{gas}}$ and $\rho_{\text{gas}}$ are temperature and mass density of the gas, respectively. In the simulation, the cool gas experiences two density regimes, below and above a number density threshold of $0.13~{\text{cm}}^{-3}$. Gas with a density above this threshold is considered star-forming. We assume that dust can survive in these cool diffuse regions, and in the dense star-forming domains below the threshold line (\ref{T12}). 

For each of these eligible gas cells the dust density $\rho_{\text{dust}}$ is calculated from 
\begin{equation}
\rho_{\text{dust}} = f_{\text{dust}}\, Z_{\text{gas}}\, \rho_{\text{gas}},
\label{rhodust}
\end{equation}
where $Z_{\text{gas}}$ is the gas metallicity. The parameter $f_{\text{dust}}$ represents the amount of metallic gas confined in dust grains and this is the second free parameter in our procedure. 
As indicated in the next section, we keep $f_{\text{dust}}$ constant as it is difficult to constrain this parameter from observational studies. 
The assumed dust model is the THEMIS model for the diffuse ISM \citep{Jones2017}, with 15 grain size bins \citep{Kapoor2021}.
In a recent study, \citet{Camps2022} analysed differences between the THEMIS and the commonly used \citet{Zubko2004} dust model in the context of the post-processing of the cosmological galaxy simulations. 
They state in their Appendix A, that since the THEMIS model is generally more emissive and that its grain population shows up to 25 per cent more extinction than \citet{Zubko2004}, the agreement with the observations is wavelength dependant. 
We decided to use the THEMIS model as it is physically motivated (it is based on the laboratory data) and it reproduces observed properties of the Galactic dust \citep{Jones2017}.

Previous studies applied the same recipe to derive the diffuse dust mass, all including the star-forming gas, however, with a different temperature limit for the diffuse gas. For example \citet{Camps2016, Camps2018},  \citet{Trayford2017}, and \citet{Vogelsberger2020b} included only non star-forming gas with $T_{\text{gas}} < 8000~{\text{K}}$, \citet{Schulz2020} and \citet{Popping2022} applied $T_{\text{gas}} < 75000~{\text{K}}$, \citet{Ma2019} adopted $T_{\text{gas}} < 10^6~{\text{K}}$ while \citet{Rodriguez-Gomez2019} did not include diffuse gas at all. Recently, \citet{Hayward2021} employed the threshold line (\ref{T12}) to calculate submm fluxes for the TNG galaxies (without performing RT simulations), and \citet{Kapoor2021} applied it in post-processing of the Auriga simulations \citep{Grand2017}. 

\begin{figure}
 	\includegraphics[width=\columnwidth]{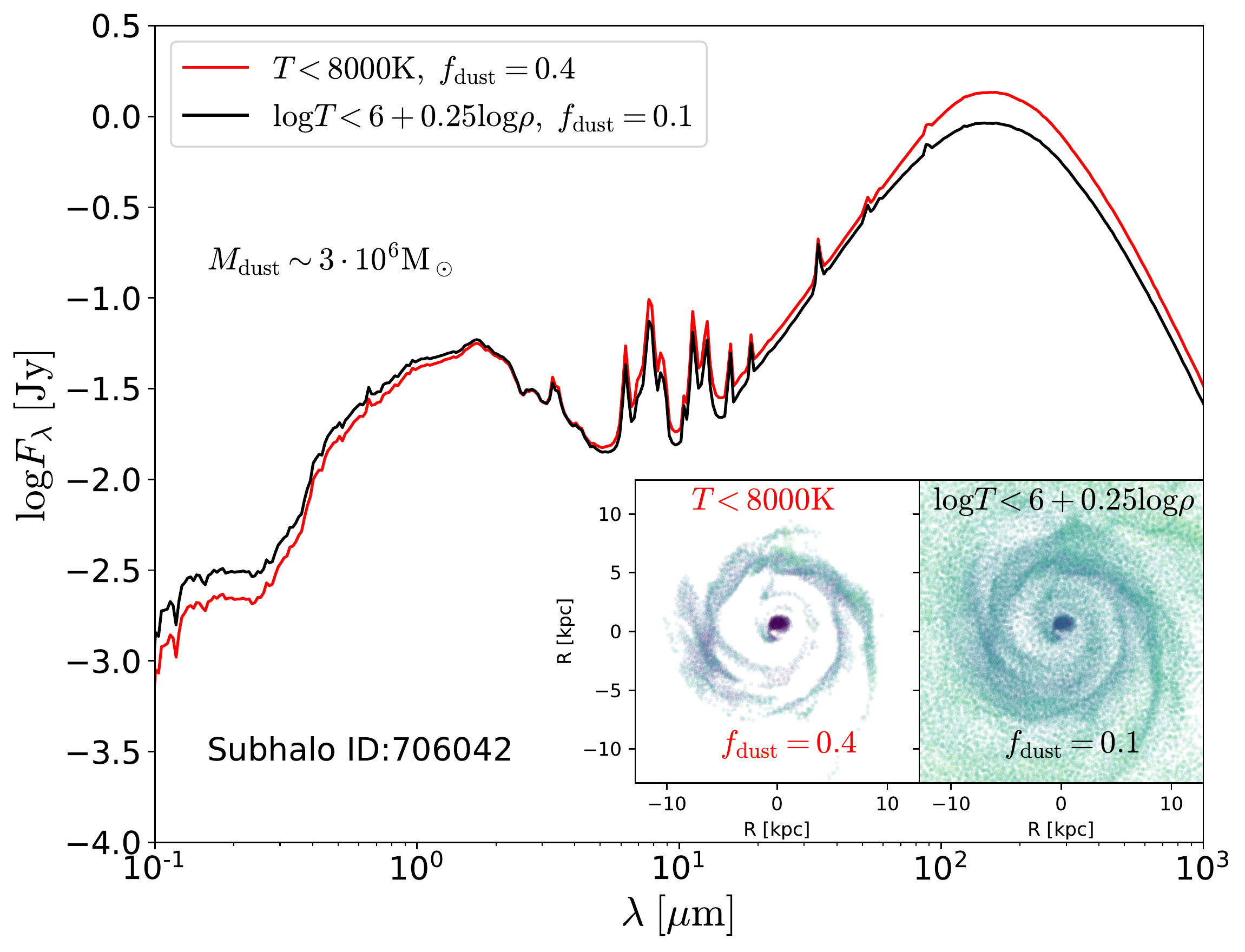}
     \caption{Two SEDs of the same TNG50 galaxy representing the effect of applying two different gas thresholds in the RT modeling. The dust mass is the same. The inset plots show the dust mass distribution for the two recipes.}
    \label{fig:sedins}
\end{figure}

It is a challenging task to constrain this threshold as the processes involving dust production, destruction and transport are still poorly understood. However, we note that the change of the cut can considerably increase the number of dust cells, and potentially significantly affect the galaxy attenuation curve. This would in turn affect the galaxy SED as it would modify the star-dust geometry \citep[e.g.,][]{Narayanan2018b}, even if the amount of dust does not change (by regulating $f_{\mathrm{dust}}$). We show this effect in Fig.~{\ref{fig:sedins}}, where the two curves represent the SED of the same galaxy, with the same dust mass but with the different dust mass distribution (shown in the inset), based on the applied cut. We employ two distinct $f_{\mathrm{dust}}$ to scale the dust mass, and the properties of the SF regions are the same. We can see that the SED corresponding to the model with a compact dust distribution (red line) is more attenuated at short wavelengths and has a higher far-infrared (FIR) output. In the other case, the same dust mass has a more diffuse distribution, causing more star light to escape.
The dust masses from the \textsc{cigale} (see Sec. \ref{subsec:cigale}) fitting of these two SEDs, as expected, are not the same.
The flux in FUV changes 0.15 dex, while around the peak of the IR emission ($\sim \um{200}$) it differs 0.17 dex.

Throughout this paper, we adopt the \cite{Torrey2012} recipe since it removes the hot and diffuse gas and, at the same time, incorporates all SF gas. 
Furthermore, \citet{Kapoor2021} found a slightly better agreement for the spatially resolved morphology parameters for the Auriga simulations with this dust recipe, compared to the more restrictive recipe from \citet{Camps2016}.

\subsubsection{Other RT parameters}
In the TNG simulations, the gas hydrodynamics is modelled on a Voronoi mesh, which we also use to discretize the gas distribution as well as construct our medium (dust) system. However, for the dust discretization, we apply an octree grid to increase performance since the native Voronoi grid is not necessarily the most optimal grid for the RT simulations in terms of resolution \citep{Camps2013, Kapoor2021}. The octree grid implementation requires fixing several parameters. The maximum and minimum levels of grid refinement are important to ensure the balance between the resolution and the efficiency of RT procedure. 
The maximum level is set to 12, based on the aperture size of the biggest galaxy ($10R_{1/2,{\text{max}}}\approx 170$~kpc) and the gravitational softening length of the TNG50 simulation ($\epsilon =0.074$ kpc). 
For the minimum level, our test converged to 6. An additional and less strict condition is that the grid will refine as long as the maximum dust mass fraction is above $2\times 10^{-6}$. Finally, the selected number of photon packets to launch is $5\times 10^7$.

To verify the robustness of our procedure, as presented in Appendix \ref{sec:A-tests}, we performed several tests including the derivations of the relative error, the variance of the variance and the Monte Carlo noise.

For each galaxy in our sample, we calculate the fluxes for three different angles: face-on, edge-on, and random. The first two are derived based on the galaxy stellar angular momentum and the third is given by the galaxy orientation in the simulation. Unless stated otherwise, for our analysis we exploit the random orientation.

Finally, \textsc{skirt} 9 allows for a high flexibility regarding the use of wavelength grids. Contrary to previous versions of the code, where a single fixed wavelength grid was used throughout the simulation, \textsc{skirt} 9 uses multiple independent wavelength grids, each specialized for a particular purpose \citep{Camps2020}. 
To calculate the dust emission spectrum in each spatial cell, we apply a nested grid with 100 logarithmically spaced bins, in the range from 0.2 to \um{2000}, and 200 additional bins in the range from 3 to \um{25}. 
In this way the emission region of small, stochastically heated dust particles is densely covered.
The radiation field wavelength grid, used for storing photon packet contributions, contains 25 bins in range from 0.02 to \um{10}. The total galaxy SED is collected in 450 bins between 0.02 and to \um{2000}. 
These choices are motivated by a tradeoff between accuracy and performance \citep{Camps2020}.

\subsection{The RT Calibration}
\label{subsec:cal}

The RT procedure introduced in the previous subsection has two global free parameters: the dust-to-metal fraction $f_{\text{dust}}$ and the PDR clearing timescale $\tau$. In this subsection we discuss how these two parameters are determined using a comparison to observations. 

\subsubsection{The DustPedia observational sample}
\label{subsubsec:DP} 

To accurately calibrate the RT procedure of the simulated galaxies, it is necessary to have an appropriate observational sample. 
For this purpose, we decided to use the DustPedia sample \citep{Davies2017} of 814 local galaxies with \textit{Herschel} and WISE 3.4 $\mu m$ detections. The photometry of the sample is derived using the matching aperture method\footnote{For every galaxy the flux in every band is calculated inside the same characteristic ellipse.} over a wide wavelength range, from the FUV to the submm \citep{Clark2018}. The DustPedia sample covers different environments \citep{Davies2019} and contains sizeable samples of galaxies with different morphologies, stellar masses and dust masses \citep{Bianchi2018, Nersesian2019, Mosenkov2019}. Therefore, it is a representative sample of the local Universe with data covering the entire UV–submm wavelength range, optimal for our purpose.

\subsubsection{Selection of a TNG50 calibration sample}

\begin{figure}
 	\includegraphics[width=\columnwidth]{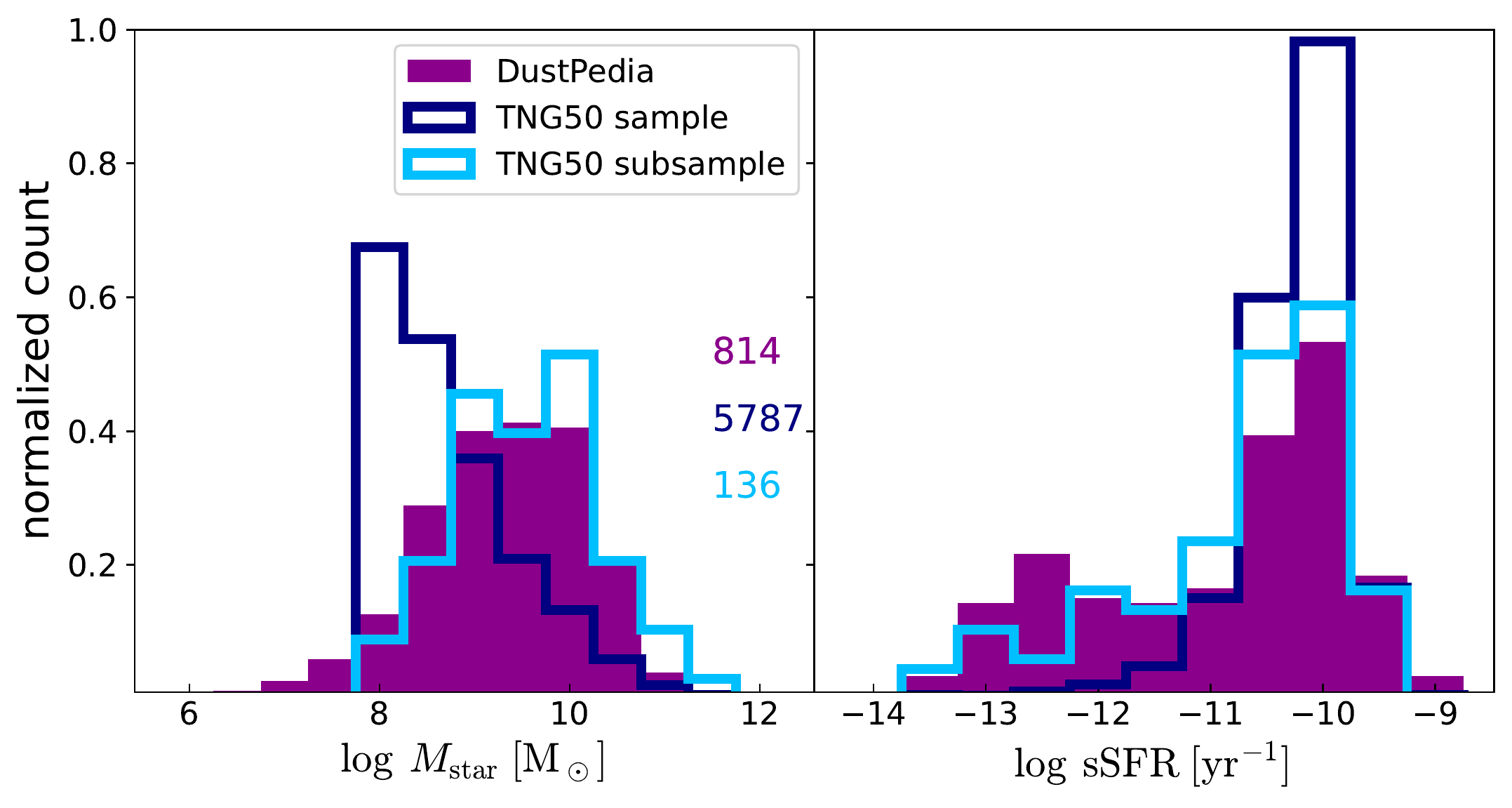}
     \caption{Distributions of stellar mass (left) and sSFR (right), for the DustPedia observed sample (dark magenta), the star-forming TNG50 simulated galaxies (navy blue) and the TNG50 subsample chosen to match the DustPedia sample and used for the RT calibration (light blue). Numbers represent the number of galaxies in each sample.}
    \label{fig:hist}
\end{figure}

The complete sample of the TNG50 galaxies we focus on in this study contains more than 14000 galaxies, as mentioned in Sec.~{\ref{sec:TNG}}.
Since the calibration of the RT procedure of a sample this large would be immensely expensive, we carefully select a smaller calibration subsample.
Since we will compare the simulations' calibration sample with the DustPedia observational sample, it is important that these are similar. To achieve that, first we select TNG50 galaxies at redshift $z=0$, as the DustPedia sample comprises only local galaxies. 
We leave out galaxies with no SF (navy blue triangles in Fig. \ref{fig:select}), as all DustPedia galaxies have non-zero SFR, reducing the TNG50 sample to 5787 galaxies.
Fig.~{\ref{fig:hist}} compares the distributions of stellar mass and sSFR of the two samples, DustPedia and TNG50 in dark magenta and navy blue, respectively. 

The stellar masses and sSFR values of DustPedia galaxies are derived using the \textsc{cigale}
SED fitting tool (see Sec. \ref{subsec:cigale}), and for TNG50, we use the intrinsic data (inside $2R_{1/2}$). We then create a matched TNG50 subsample by random (rejection) sampling from the DustPedia galaxy stellar mass and sSFR distributions. 
For both distribution we identify 20 bins, and then we draw the appropriate number of galaxies in each bin.
After the stellar mass selection, the initial TNG50 sample drops to 1730 galaxies, from which we try to match the sSFR distribution, so the final sample has 136 galaxies, shown in cyan. 
This new TNG50 sample, our calibration sample, is of manageable size and is more similar to the observed one, at least based on stellar mass and sSFR.

\subsubsection{Aperture}
\label{subsec:aper}
\begin{figure}
	\includegraphics[width=\columnwidth]{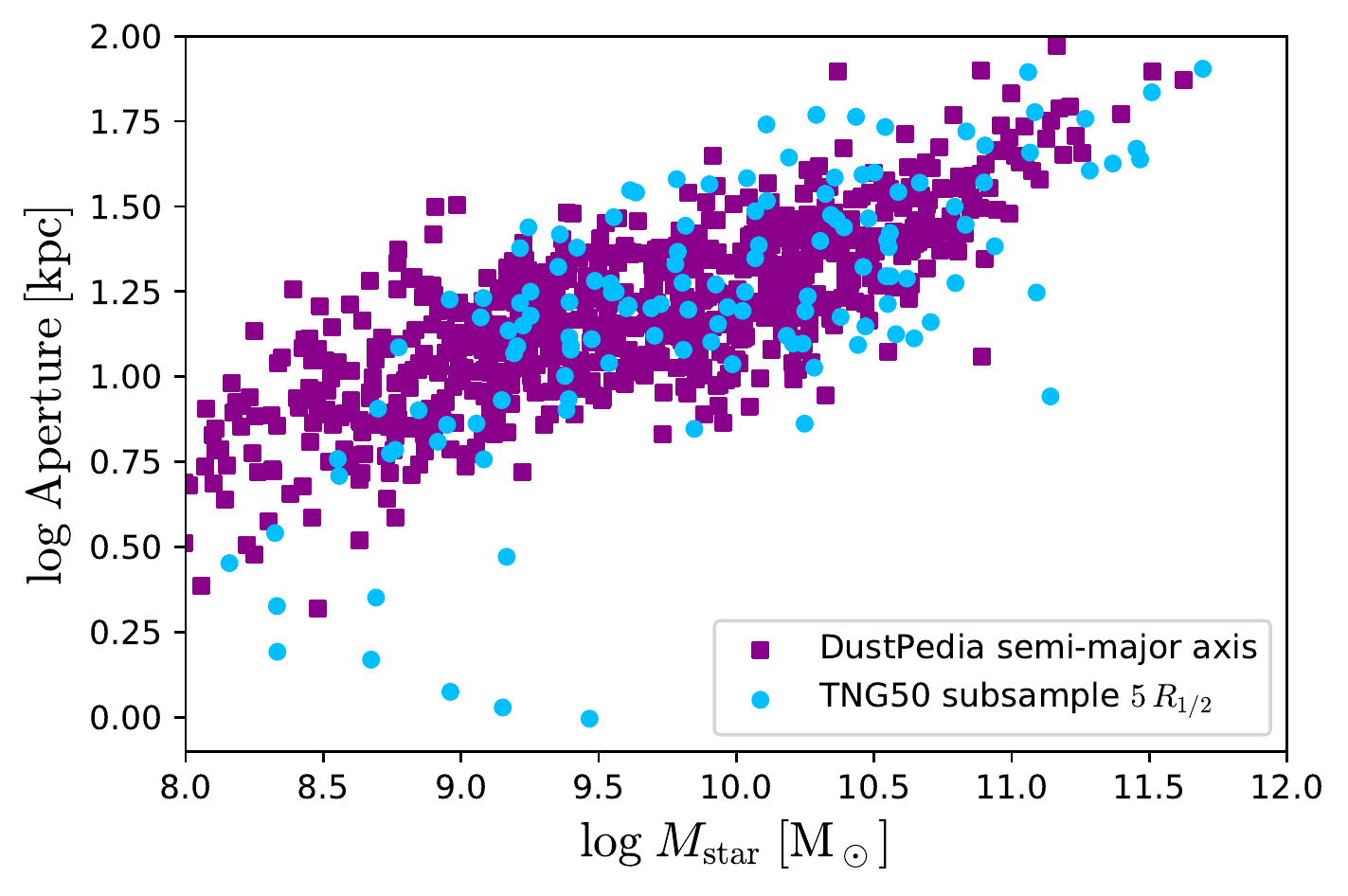}
     \caption{DustPedia apertures as a function of stellar mass (dark magenta), compared to the $5\,R_{1/2}$ apertures as a function of stellar mass for the TNG50 calibration sample (cyan). Here, the stellar masses for the DustPedia galaxies are inferred using \textsc{cigale}, those for the TNG50 galaxies are measured by summing the masses of the individual stellar particles.}
    \label{fig:aper_mstar}
\end{figure}

The next step is to define the galaxy size or rather the aperture confining most of the galaxy light in different bands for each simulated galaxy. Different observational studies adopt different aperture definitions \citep[e.g.][]{Graham2005,Hill2011} and finding their analogues in the simulation realm is not straightforward. To tackle this problem, previous studies adopted various aperture definitions for the simulated galaxies: accounting for all particles (cells) gravitationally bound to the galaxy \citep{Torrey2015}, multiplying $R_{1/2}$ by a specific factor \citep{Rodriguez-Gomez2019, Schulz2020, Popping2022}, or applying a constant aperture for all galaxies \citep[e.g.][]{Camps2016, Camps2018, Trayford2017, Vogelsberger2020b}. 
To achieve the most consistency, we decided to mimic the aperture size distribution of the DustPedia sample as a function of stellar mass. Fig.~\ref{fig:aper_mstar} represents this attempt. We see that the TNG50 3D apertures of $5\,R_{1/2}$ correspond well to the DustPedia apertures\footnote{We assume the semi-major axis of the master elliptical aperture as the DustPedia aperture.} on the whole stellar mass dynamic range. How this aperture change affects the sample selection procedure is discussed in Appendix \ref{sec:A-aper}.
If not otherwise stated, the aperture of $5\,R_{1/2}$ is fiducial in this study, however, in Sec. \ref{subsubsec:LF-aper} we will discuss the effects of aperture on the LFs.

\subsubsection{SED fitting with \textsc{cigale}}
\label{subsec:cigale}

Although each DustPedia galaxy on average contains reliable fluxes in 20 broadband filters, we perform an SED fitting procedure to improve the SED coverage and infer physical properties. Furthermore, to provide the most consistent comparison of the observational and the mock samples, we perform the same SED fitting procedure on the TNG50 calibration sample.
We also use the same set of bands as those used for the fitting of the DustPedia galaxies. 
We exploit the SED fitting tool \textsc{cigale} \citep{Boquien2019}, which relies on the Bayesian inference technique and which employs a range of highly configurable models. For our purpose, we adopted a set of parameters that were fine-tuned for the DustPedia sample and are described in detail in \citet{Nersesian2019}. In summary, we assume a truncated delayed SF history \citep{Ciesla2016}, the \citet{Bruzual2003} SSP libraries, the THEMIS model for the dust emission \citep{Jones2017}, and a modified \citet{Calzetti2000} curve \citep{Boquien2019} for the dust attenuation. Compared to \citet{Nersesian2019}, the only parameter we updated in this study is the IMF, where we selected \citet{Chabrier2003} instead of \citet{Salpeter1955}, consistent with the cosmological and RT simulations.

The \textsc{cigale} fitted SEDs differ from the ones derived from \textsc{skirt} always less than 0.1 dex and mostly less than 0.05 dex.
Therefore, we are confident that the fitted and the input fluxes can be used interchangeably. 
Furthermore, in Appendix {\ref{sec:A-tng-cig}} we calculate the difference between the intrinsic galaxy properties and those derived from \textsc{cigale}. 
The median deviation is less than 0.14 dex. 
Therefore, for the calibration we will use the physical properties and fluxes derived from the SED fitting for both observational and simulated 'calibration' samples. In this manner, both samples have the same wavelength coverage and are treated adopting the same models, alleviating the potential biases during the calibration.

\subsubsection{Determination of the RT free parameters}

\label{subsubsec:calib}
The essential and final step before we run the RT procedure on the whole TNG50 sample is to specify the remaining free parameters of the procedure: $\tau$ and $f_{\mathrm{dust}}$.

\begin{figure}
	\includegraphics[width=\columnwidth]{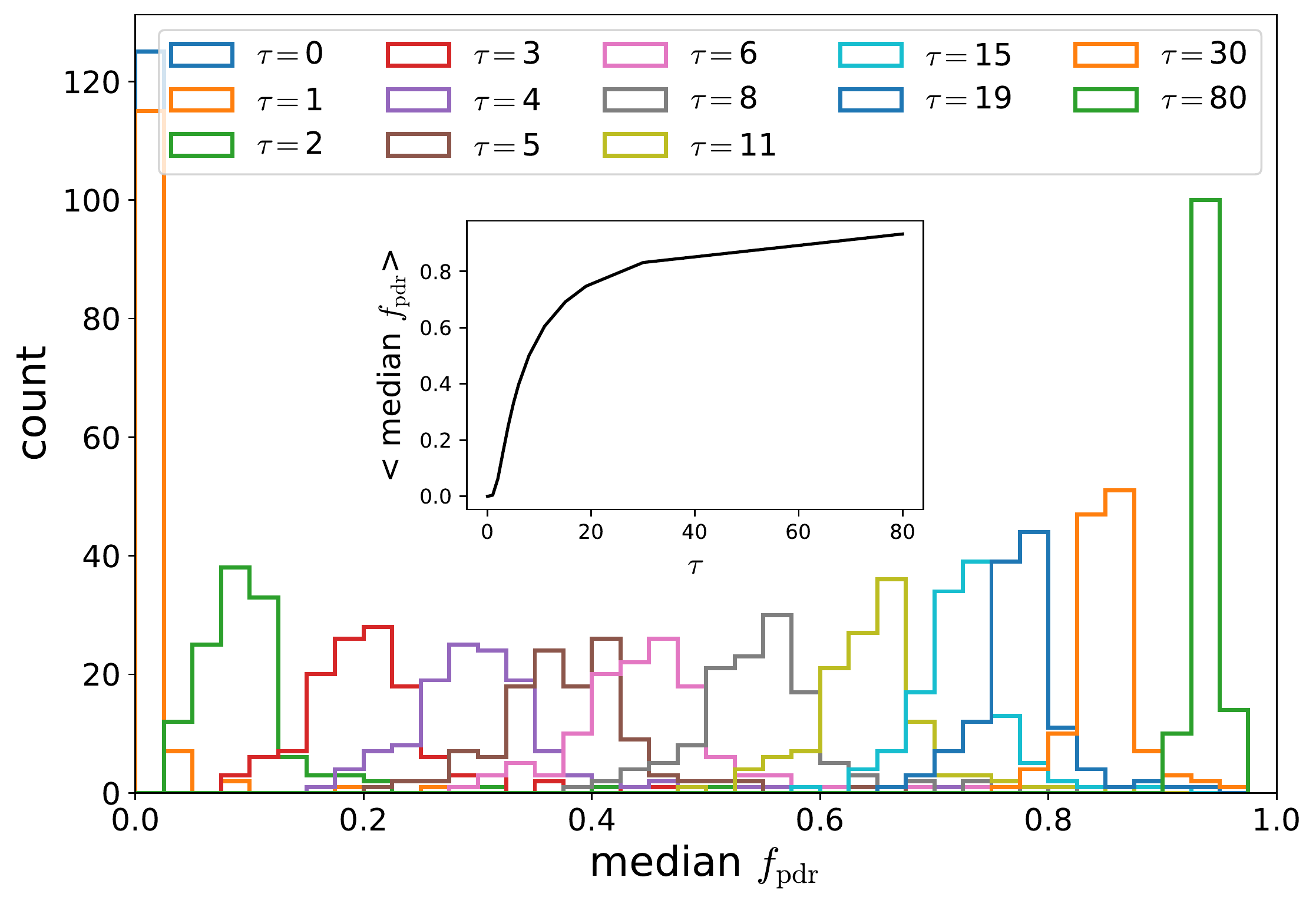}
     \caption{Distributions of the median $f_{\mathrm{PDR}}$ for different values of $\tau \mathrm{[Myr]}$, for the 136 TNG50 galaxies in the calibration subsample. 
     The inset shows that with the increase of $\tau$, distributions of the median $f_{\mathrm{PDR}}$ peak at higher values.}
    \label{fig:fpdr_tau}
\end{figure}

The first parameter ($\tau$) controls the level of exposure of \textsc{hii} regions, and ranges from 0 to $\infty$.  The higher this value, the more the H\textsc{ii} region is covered by the PDR layer. 
By construction, this parameter also affects the diffuse dust, as the amount of radiation that does not illuminate the dust in SF regions (if $\tau$ is small) will propagate into the ISM to heat the diffuse dust.
Contrary, if $\tau$ is large, the energy will not leave the SF region, and the heating of the diffuse dust will be reserved mostly for the older stars. 
We chose to vary the parameter $\tau$ instead of $f_{\text{PDR}}$ (defined in formula \ref{eq:tau}), so we can mimic the spread in the covering factor relating to the clearing of SF regions as time progresses. Fig.~\ref{fig:fpdr_tau} shows how the median $f_{\mathrm{PDR}}$ of our TNG50 subsample changes with $\tau$. 
For very low $\tau$ all galaxies have almost all SF regions completely open. 
At values as high as $\tau=11$~Myr, most of the SF regions are $60\%$ covered. 

The dust-to-metal fraction ($f_{\mathrm{dust}}$) directly controls the diffuse dust mass, as explained in Sec. \ref{subsec:gal_part}. 
A growing body of literature has investigated this parameter for various galaxies and at different redshifts. While some studies suggest a trend with different galaxy properties like metallicity, stellar mass, or redshift \citep{DeCia2013, Zafar2013, RemyRuyer2014, DeVis2019, Peroux2020}, these relations all have a relatively wide spread. \citet{DeVis2019} reported that for a DustPedia sub-sample of 364 galaxies\footnote{Only for 364 galaxies in the DustPedia sample the necessary data to calculate $f_{\mathrm{dust}}$ are available.} the median $f_{\mathrm{dust}}$ is 0.2 with a range [0.03-0.64]. 
However, the physical properties in the observational studies are derived in a different way than in the simulations: it is therefore challenging to constrain this parameter directly from observations \citep[although see][]{Popping2022}. 
Additionally, our current study focuses on galaxies with stellar masses above $10^8~{\text{M}}_\odot$ and at low redshift, thus the potential evolution of $f_{\text{dust}}$ would be less supported than at higher redshift. 
Accordingly, we opt not to vary $f_{\text{dust}}$ for different galaxies, and to treat it as a global free parameter. We explored a recipe variation where we add random noise to $f_{\text{dust}}$ for different dust cells, while keeping the integral value constant. The results are presented in Appendix {\ref{sec:A-tests}} and the deviations are negligible.

In their work on the EAGLE simulations, \citet{Camps2016} calibrated the RT procedure by comparing mock and observational galaxy scaling relations. They used three scaling relations that incorporated several bands: three SPIRE, three SDSS and the GALEX NUV band. In a later study, \citet{Trcka2020} speculate that the exclusion of the FUV and MIR bands during the calibration can explain the high UV output seen for the EAGLE simulations \citep{Baes2019}, as the emission from the SF regions was not constrained sufficiently. Therefore, we decided to extend the SED coverage used in the calibration by including FUV, WISE 22, MIPS 70 and PACS 160 bands. The set of scaling relations includes those from \citet{Camps2016} (with $L_{\um{3.4}}$ and $L_{\um{250}}$ instead of stellar and dust mass, respectively), one with the alternative sSFR proxy ($L_{FUV}/L_{\um{3.4}}$ instead of $\mathrm{NUV}-r$) and four showing extensive relations. With this set, we incorporate the data from 10 different bands, from FUV to submm. As stated previously, all quantities are derived with the \textsc{cigale} tool.

\begin{figure*}
	\includegraphics[width=1.84\columnwidth,keepaspectratio]{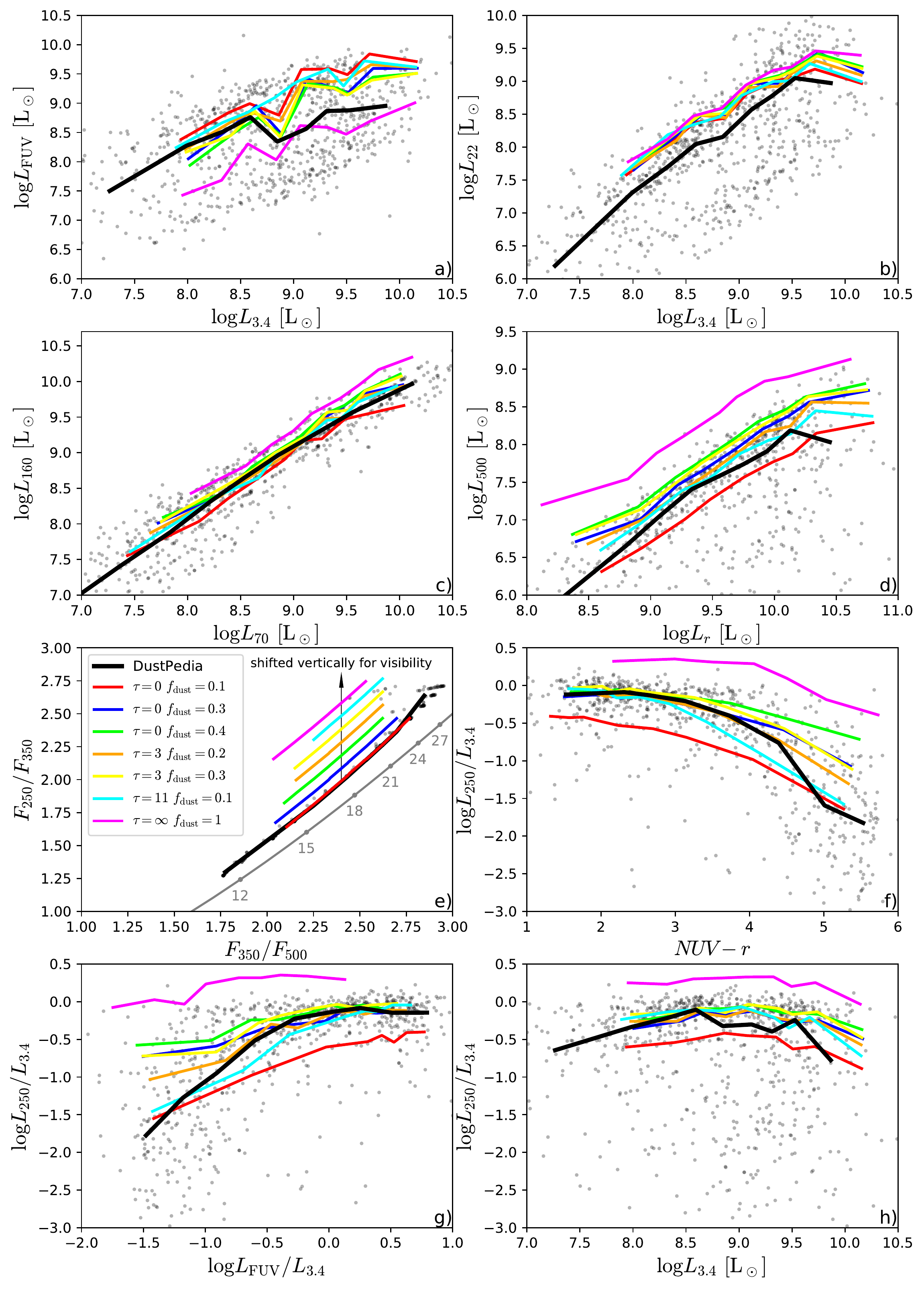}
     \caption{
     Eight $z=0$ galaxy scaling relations derived from fitting galaxy SEDs with CIGALE for both observed (DustPedia, black solid thick curves) and simulated (TNG50, coloured solid thin curves) galaxies. For the latter, curves represent the running medians of the different SKIRT RT recipes. The grey line in plot e) represents a modified blackbody (MBB) curve assuming the THEMIS dust model. The grey numbers correspond to the dust temperature in K. If not shifted, the lines in plot e) would overlap. The data points correspond to the DustPedia galaxies.}
    \label{fig:LLLLLcig}
\end{figure*}

The results of our calibration exercise are shown in Fig.~{\ref{fig:LLLLLcig}}. The DustPedia data are shown in black. For the visual comparison of different recipes, we show the running median lines. Moreover, for a more quantitative valuation of the recipes, we performed the two-dimensional Kolmogorov-Smirnov test \citep[KS test,][]{Kolmogorov1933, Smirnov1948, Peacock1983, Fasano1987} on each scaling relation. 
This test provides us with two statistics: $p$-value and $D$-distance between the distributions.
For each recipe, we derive the median\footnote{Using the mean value instead of the median does not affect the results.} of the eight $D$-values and compare these to find the recipe with the smallest distance from the DustPedia sample.  
First, we ran a set of RT simulations with extreme parameter values to explore the range of the parameter space. 
These are shown in Fig.~{\ref{fig:LLLLLcig}} as red and magenta median lines. 
Comparing to the DustPedia in black, the two generally show poor agreement: the KS test gives $D=0.32$ and $D=0.48$ for the prescription with $\tau=0$ and $f_{\text{dust}}=0.1$, and $\tau=\infty$ and $f_{\text{dust}}=1$, respectively, i.e., with minimum and maximum dust in both compact and diffuse regimes. These differences emphasise the importance of dust in the galaxy modelling. 

\begin{figure*}
	\includegraphics[width=2\columnwidth]{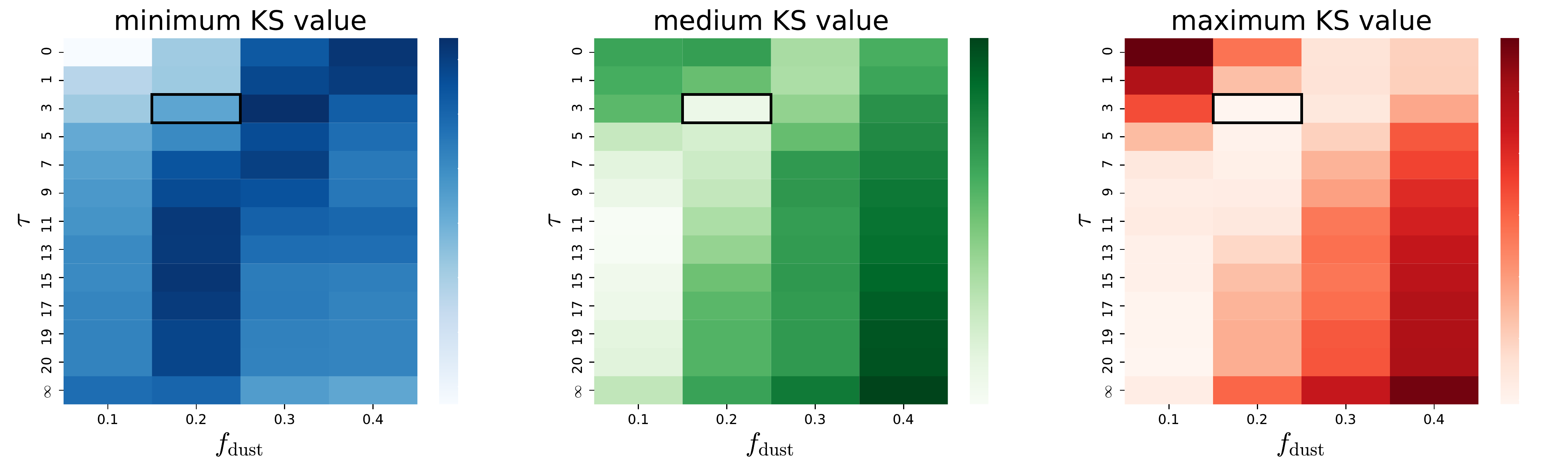}
     \caption{Minimum (left), median (centre), and maximum (right) KS statistic $D$ of the eight relations shown in Fig.~\ref{fig:LLLLLcig}, for the different values of $f_{\text{dust}}$ (x-axis) and $\tau$ (y-axis). Lower values that indicate smaller distance between the distributions are mostly located on a diagonal strip. For our fiducial RT modeling we adopt $f_{\text{dust}}=0.2$ (x-axis) and $\tau=3$ (y-axis), indicated with black rectangles.}
    \label{fig:KScig_hm}
\end{figure*}

\begin{figure*}
	\includegraphics[width=2\columnwidth]{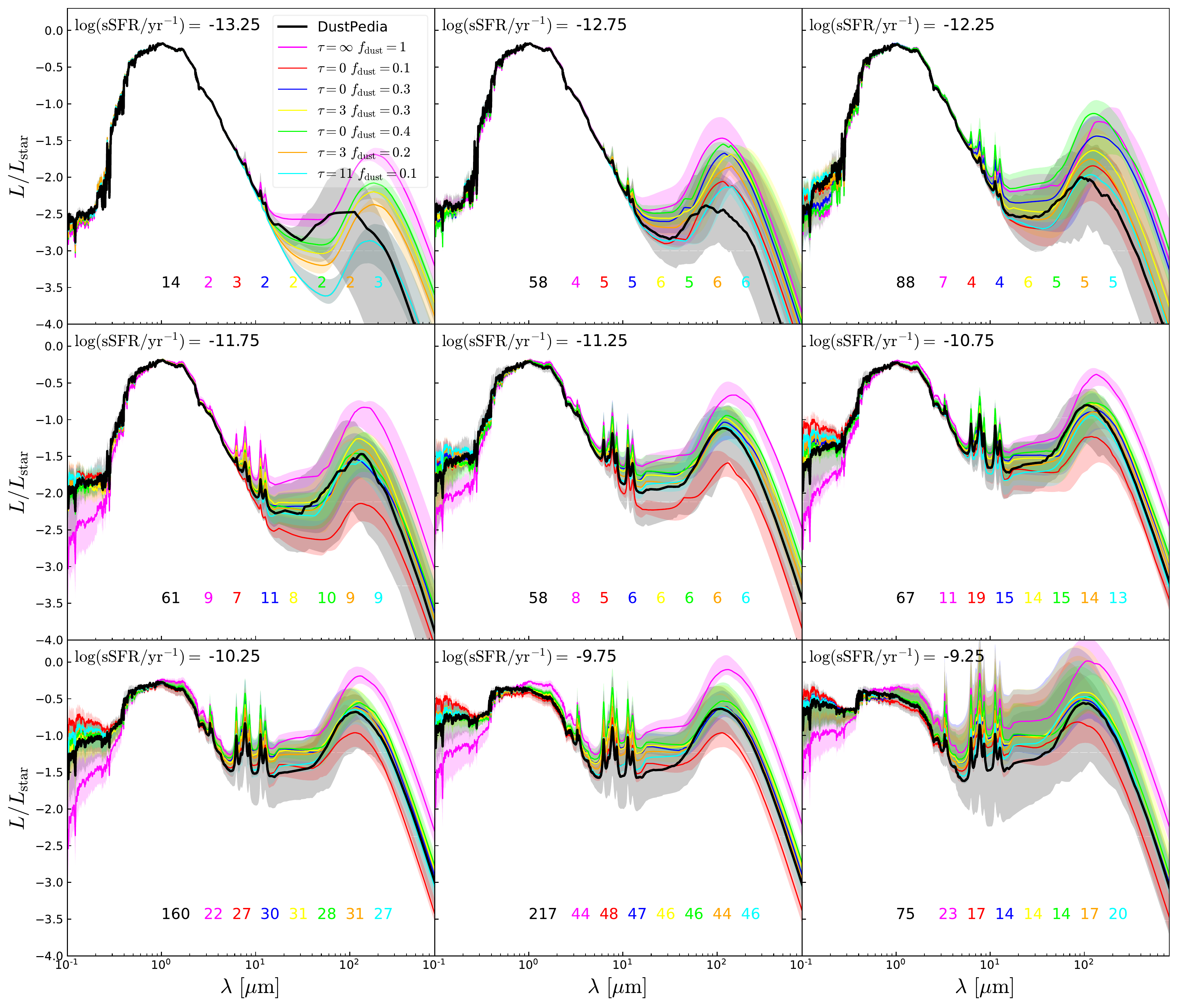}
     \caption{Median SEDs for the DustPedia observed galaxies (black) and for the TNG50 subsample for a number of RT-modeling choices, in bins of $\log$ sSFR at z=0. The shaded regions represent 16$\%$ - 84$\%$. The numbers of galaxies in each bin and each recipe are also shown, at the bottom.}
    \label{fig:seds_cig_ssfr}
\end{figure*}

Beside these two extreme recipes, we explored a full range of other, more physical options. The results of the KS tests for some of these are shown in Fig.~{\ref{fig:KScig_hm}}. These panels show that there are several favourable parameter combinations, mostly towards low $f_{\text{dust}}$ and high $\tau$ values. Some of them are shown as well in Fig.~{\ref{fig:LLLLLcig}}.

While the results of the KS test show that there is a degeneracy among the two parameters (increasing $\tau$ and decreasing $f_{\text{dust}}$, and the opposite give a constant $D$ value), such degeneracy can be broken if we analyse the median lines of the individual scaling relations. 
\begin{itemize}
\item 
Panel \ref{fig:LLLLLcig}h represents the relation between the luminosity at $\um{3.4}$ and $L_{250}/L_{3.4}$ (stellar and specific dust mass proxy, respectively). While the recipes in colour show similar results over the wide dynamic range, for the most massive systems the dust mass trends depend mostly on $f_{\text{dust}}$. This is expected since most massive systems have little or no SF regions.
\item
On the other hand, $L_{3.4}$, which traces emission from evolved stars, remains fairly constant with the recipe change. This is foreseen since we are not modifying this source, and the dust extinction is weak.  However, hot dust also emits at these wavelengths. This can be seen in the same plot as the low $L_{3.4}$ limits of the medians change depending on the amount of diffuse dust (depending on $f_{\text{dust}}$) and the capability of the young stars to heat the dust (depending on $\tau$).
In contrast, for the massive (and passive, as they have lower $L_{250}/L_{3.4}$) end this effect would be insignificant \citep{Norris2014}. 
\item
Another relation that can be used to untangle the effects of the two parameters is in panel~\ref{fig:LLLLLcig}g, or equivalently panel~\ref{fig:LLLLLcig}f. On the x-axis they have a similar tracer for the sSFR, while on the y-axis they show a proxy for the specific dust mass ($M_{\text{dust}}/M_\star$). For galaxies with low sSFR, the parameter $\tau$ becomes irrelevant (recipe with a different $\tau$ and the same $f_{\mathrm{dust}}$ converge). If we follow the median lines in the direction of increasing UV for a fixed stellar mass, the results become more sensitive to the content of the SF regions. 
\item
The plot in panel~\ref{fig:LLLLLcig}e shows the FIR colours. This relation is constrained by the SED fitting, however, the same relation with the `raw' pre-\textsc{cigale} data is also fairly tight, particularly for the simulated galaxies, as there are no observational effects \citep[see also][]{Camps2016, Camps2018, Kapoor2021}. As the medians overlap, we shift them by 0.1 dex to increase visibility. This relation gives us information about the galaxy dust temperature, as these flux ratios describe the slope of the IR emission. This color-color plot can be used to check whether the TNG50 galaxies have physical dust temperatures, since the limited resolution can produce an unrealistic stellar/dust configuration \citep{Camps2018}. 
All galaxies from the TNG50 calibration sample are sufficiently resolved and they populate the same temperature range as the DustPedia galaxies, however, mainly with intermediate values. 
As we see from the positions of the median lines, it is difficult to clearly unravel the individual contribution of the parameters, especially since the relation is also sensitive to the geometry of its components.
\item
The relation in panel~\ref{fig:LLLLLcig}c between luminosities in bands from both sides of the IR peak emission shows high agreement regardless of the recipe, while panel~\ref{fig:LLLLLcig}d shows that, on a contrary, in some cases the discrepancy can be up to an order of magnitude. Furthermore, in this plot (similarly also in panel~\ref{fig:LLLLLcig}h) we can see that at the high stellar mass end (high $L_{r}$) the dust output, while decreasing, is still too high. These two panels generally show poorer agreement irrespective of the recipe.
\item
Finally, in panel~\ref{fig:LLLLLcig}a, the break in the median line of the DustPedia galaxies demonstrates the bimodality in the relation, while this trend is less obvious for the simulated sample, and for the high stellar mass galaxies, the FUV emission is too high. 
\end{itemize}

It is clearly challenging to select one ideal recipe: there is no single recipe that scores best in each of the individual panels of Fig.~{\ref{fig:LLLLLcig}}. This is also evident from Fig.~{\ref{fig:seds_cig_ssfr}}, which shows the effect of the different recipes on the global galaxy SEDs. The individual panels contain the median SEDs of a number of recipes for different sSFR bins. The SEDs are normalised by the total stellar luminosity derived from \textsc{cigale}. 
We can see that, for low sSFR galaxies, dust emission greatly varies between the recipes, with minimal change in the UV. 
These galaxies have few or no young stars, therefore intrinsically very little UV to be attenuated. 
For higher sSFR values the UV and FIR domains are fitted remarkably well employing a number of recipes, while the MIR range shows increasingly poorer agreement.  This implies that for every parameter configuration the TNG50 galaxies with high sSFR will have an excess of very hot dust. Reviewing the different galaxy populations in Fig.~{\ref{fig:seds_cig_ssfr}}, we see that in distinct sSFR bins and wavelength regimes, different recipes show better agreement.
Even the nonphysical recipes with $f_{\mathrm{dust}}=0.1$ and $\tau=0$, and with $f_{\mathrm{dust}}=3$ and $\tau=\infty$, do not show the highest discrepancies for $\log (\mathrm{sSFRs}/\mathrm{yr})$ between $-13$ and $-12.25$. 
We do note, however, that most of the (physical) recipes presented in Figs.~{\ref{fig:LLLLLcig}} and {\ref{fig:seds_cig_ssfr}} are within the errors and spread of the DustPedia data along the whole wavelength range, meaning that our procedure is robust and that the final selection is merely fine-tuning. 

Based on the deviations between the median SEDs in different sSFR bins and the KS scores, we conclude that, on average, the recipes with $f_{\text{dust}}>0.2$ are the least favourable, regardless of $\tau$ ($D>0.282$). Furthermore, the recipes with a fixed $f_{\text{dust}}$ and a high $\tau$ show similar results, as the differences between the total covering fractions are getting smaller (see equation \ref{eq:tau} and Fig.~\ref{fig:fpdr_tau}). 

The two parameter combinations that we finally consider are the one with $f_{\text{dust}}=0.1$ and $\tau=11$~Myr ({\tt{fd1-t11}}, hereafter) with $D=0.243$, and the one with $f_{\text{dust}}=0.2$ and $\tau=3$~Myr ({\tt{fd2-t3}}, hereafter) with $D=0.252$. 
The main differences between these two recipes are in the MIR and FIR for galaxies with low sSFR (orange and cyan lines in Fig.~{\ref{fig:LLLLLcig}} and Fig.~{\ref{fig:seds_cig_ssfr}}). For all sSFR bins, the {\tt{fd2-t3}} dust prescription has a higher MIR output. This is a consequence of, on average, mostly open H\textsc{ii} regions (see Fig.~{\ref{fig:fpdr_tau}}) from which the UV radiation is now heating more diffuse dust, compared to the {\tt{fd1-t11}} scheme. Both options produce comparable FIR output for moderate and high sSFRs. For the low sSFR galaxies, there is a lack of compact dust to increase the IR emission for high values of $\tau$, therefore the discrepancy between the two dust allocation recipes is higher. Nevertheless, for the whole calibration sample, the difference in the dust mass (from \textsc{cigale}) is on average $\approx 0.15$ dex.
Since these two recipes similarly affect a galaxy SED, we decided to adopt {\tt{fd2-t3}} as the fiducial dust distribution, as this value of $f_{\text{dust}}$ is comparable with that of the DustPedia and other observational samples \citep[][]{DeVis2019, Zabel2021, Galliano2021}.
So, throughout this work, simulated galaxies are post-processed with \textsc{skirt} with $f_{\text{dust}}=0.2$ and $\tau=3$.
In Sec.~{\ref{subsubsec:LF-rec}} we address the effect of different RT choices on the LFs.

One question that arises is whether our results would change if we had chosen another set of scaling relations for the calibration. To test this, we analysed the distribution of the KS scores if we only include relations used in \citet{Camps2016}, and the results qualitatively do not change. Analysing a set of relations that all include one galaxy property on one axis, such as stellar mass ($L_{3.4}$) or dust mass ($L_{250}$), does not bring our results to convergence. This approach can be successfully applied to simulations of less diverse galaxy populations \citep{Kapoor2021}, however, for most properties, our sample spans three orders of magnitude, with a result that different recipes agree better for distinct galaxy populations (e.g., see Fig.~\ref{fig:LLLLLcig}a).

Finally, we note that, for a different definition of a dust tracer (see \ref{subsubsec:diff_dust}: Diffuse dust), an another set of parameters might be selected as the optimal set.
For example, \citet{Kapoor2021} compared two definitions: the one based on the threshold from \citet{Torrey2012} (the fiducial in this study), and the one restricting dust only to star-forming gas cells.
Compared to the former, the latter definition requires higher values of $f_{\text{dust}}$ for optimal results.

\section{Results}
\label{sec:res}

\subsection{UV to submm fluxes}
\label{subsec:public}

We apply our fiducial RT modeling to the whole TNG50 sample with $M_\star>10^8~{\text{M}}_\odot$ (see Fig.~\ref{fig:select}). Including galaxies without dust and the small sample used for calibration, the final sample contains 7375 galaxies at $z=0$ and 7302 galaxies at $z=0.1$. Furthermore, we apply the same RT procedure to TNG50-2, the TNG50 run with one step worse resolution (a factor of eight in particle mass, two in spatial resolution). 
To ensure a similar number of stellar particles per galaxy ($\sim$ 1000), for TNG50-2 we impose a stellar mass threshold of $\log M_{\mathrm{star}}/\mathrm{M_\odot}>8.6$.
Therefore, for TNG50-2 we have 3559 galaxies at $z=0$ and 3497 galaxies at $z=0.1$. 
For the redshift $z=0,$ the fluxes are calculated assuming the galaxies are at 20 Mpc from the detector, consistent with previous studies \citep{Camps2016, Trayford2017}. 

To validate our procedure, in Appendix \ref{sec:A-tng-prox} we investigate the differences between the true values of the galaxy stellar masses and SFRs and those calculated from the fluxes, based on recipes from the literature. 

Finally, we publish the data for future science projects. Tables \ref{tab:1} and \ref{tab:bands} summarise the available information. Broadband fluxes of the total galaxy emission are calculated in 53 bands, for two resolution options of the TNG50 simulation, two redshifts, four different apertures and three orientations. The same options are available for the galaxy SEDs. 

\begin{table}\centering
\caption{Structure of the published data. Resolution: different resolution runs of the TNG50 simulation. Redshift: galaxy redshift (if $z=0$, the flux is calculated at 20 Mpc). Aperture: 2D circular aperture used to extract SED. Orientation: angle at which the galaxy is observed. Component: the integrated flux at different wavelengths and in 53 broadbands. Every combination is provided.}

\label{tab:1}
\begin{tabular}{ll}\hline
\multirow{2}{*}{\textbf{Resolution}} & TNG50 \\
                         &  TNG50-2 \\\hline
\multirow{2}{*}{\textbf{Redshift}}   & 0   \\
                            & 0.1 \\\hline
\multirow{4}{*}{\textbf{Aperture}}   & $5 R_{1/2}$   \\
                            & $2 R_{1/2}$   \\
                            & 10 kpc   \\
                            & 30 kpc \\\hline
\multirow{3}{*}{\textbf{Orientation}}   & face-on   \\
                            & edge-on  \\
                            & random  \\\hline
\multirow{2}{*}{\textbf{Component}} & total   \\
                              & total - broadbands \\\hline

\end{tabular}
\end{table}


\subsection{Galaxy luminosity functions at \texorpdfstring{$z\le 0.1$}{z<=0.1} }
\label{subsec:LF}

The complete low-redshift TNG50 sample comprises 14677 galaxies, which allows us to inspect the LFs and compare them with the observed ones. 
We perform the study over a wide range of wavelengths, from the UV to the submm range. 
However, we omitted the MIR bands, since the available data from the literature is typically presented in a wider redshift range, which would produce an inconsistent comparison \citep{Baes2020}. 
Instead, we also investigate the total IR luminosity.

All TNG50 LFs are derived by splitting the population of galaxies in $\log L$ bins and dividing by the simulation co-moving volume, while the errors are derived following Poisson statistics. 
As we consider galaxies above a certain stellar mass threshold ($\log M_{\mathrm{star}}/M_\odot>8$ for TNG50 and $\log M_{\mathrm{star}}/M_\odot>8.6$ for TNG50-2), the low-luminosity bins are incomplete. 
To derive the minimum luminosity where these bins may be complete, in the narrow stellar mass mass bin at the low limit, we calculate the 90th percentile of the luminosity distribution. 
We select this value (calculated for each band) as a low cut for the galaxy LF.

Finally, to calculate the total IR luminosity, we implement the five-band formula for nearby galaxies from \citet{Galametz2013}, i.e.
\begin{multline}
L_{\text{TIR}} = 2.023\,L_{24} + 0.523\,L_{70} \\+ 0.39\,L_{100} + 0.577 \,L_{160} + 0.721\,L_{250}.
\end{multline}
When this formula was applied to the sample of galaxies from the EAGLE simulations \citep{Schaye2015, Crain2015} and compared to the luminosities derived from \textsc{cigale}, the results showed excellent agreement \citep{Baes2020}.

\subsubsection{Observational LFs}
\label{subsubsec:obs_lf}

In this study, we compare the simulated TNG50 LFs with a number of observational LFs. In the UV regime, in two GALEX \citep{Morrissey2007} bands, we compare to the data from the GALEX MIS survey \citep[][ $0.07<z<0.13$]{Budavari2005}, and from the shallower AIS survey \citep[][$z<0.1$]{Wyder2005}. Additionally, we use LFs covering the entire UV to NIR range \cite[][$0.013<z<0.1$]{Driver2012} from the GAMA survey \citep{Driver2011, Liske2015}. Focusing solely on the optical, \citet{Loveday2012} calculated LFs from the full GAMA survey, but with the original SDSS \citep{York2000} magnitudes. Combining SDSS, UKIDSS LAS \citep{Lawrence2007} and the MGC redshift survey \citep{Liske2003}, \citet{Hill2010} derived LFs for $0.0033<z<0.1$. In the IR domain, at $\um{250}$ we have the LF \citep[][$z<0.1$]{Dunne2011} from the H-ATLAS survey \citep{Eales2010, Rigby2011}, with redshifts from the GAMA survey.
At $\um{350}$, we compare our results with the LF from \citep[][$z\lll 0.1$]{Negrello2013} based on the \textit{Planck} ERCSC \citep{PlanckCollab2011} data. For all SPIRE bands and the total IR, we use the LFs from \citep[][$0.02<z<0.1$]{Marchetti2016} based on the HerMES survey \citep{Oliver2012}, combined with the \textit{Spitzer} Data Fusion database \citep{Vaccari2015}. 

All data from the literature are converted to AB magnitudes and to $h=0.6774$, to match the simulated data.

\subsubsection{LFs of the TNG50 simulations}
\label{subsubsec:fidu_lf}

\begin{figure*}
	\includegraphics[width=1.9\columnwidth,keepaspectratio]{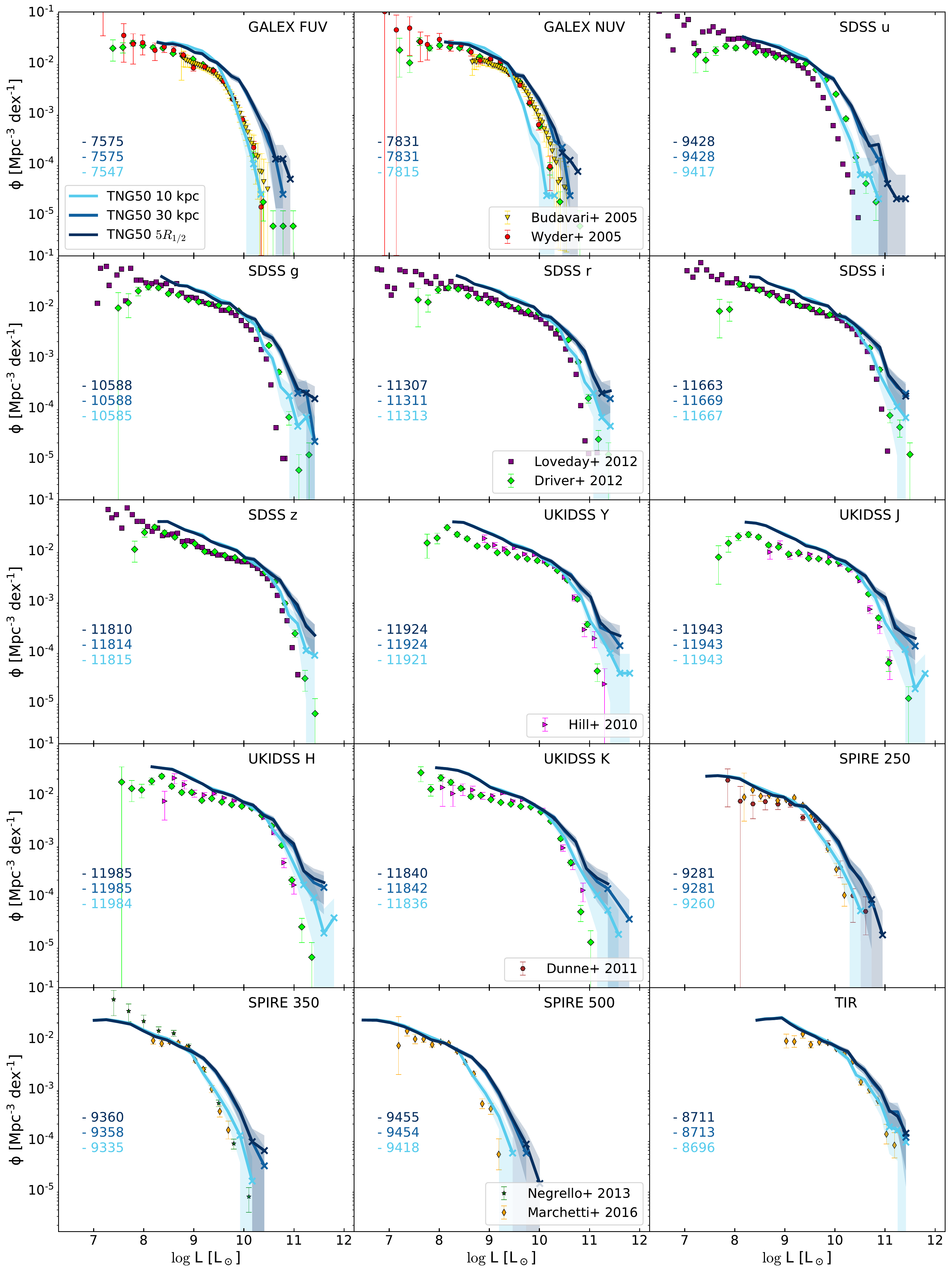}
     \caption{
     LFs in in 14 bands and the total IR LF, at $z\le 0.1$. The navy blue lines correspond to TNG50, post processed with SKIRT according to the RT procedure described and devised in the previous Sections, with light summed up from an aperture of $5\,R_{1/2}$ for each galaxy. The other blue lines show results for different apertures. The ’x’ marker depicts luminosity bins with fewer than 10 objects. Shaded region shows the Poisson error. The number on the left represents the total number of the TNG50 galaxies. We give the LFs from the simulated galaxies only above the completeness limit imposed by the selection: stellar mass $>10^{8}\; \mathrm{M_\odot}$. The coloured symbols represent observational data.}
    \label{fig:LF_fidu}
\end{figure*}

In this section, we assume the fiducial set of parameters: $5\,R_{1/2}$ apertures, random galaxy orientations, the {\tt{fd2-t3}} RT fiducial procedure recipe, the TNG50 simulation at available resolution.
Namely, we integrate the light coming from (approximately) cylindrical volumes around each galaxy with radius of $5\,R_{1/2}$ in the plane of the sky and with a depth of $10\,R_{1/2}$.
The results are shown in Fig.~{\ref{fig:LF_fidu}}.

By inspecting the fiducial TNG50+RT model (in navy blue) over a range of wavelengths, we immediately notice that the TNG50 LFs tend to overpredict the observational ones.
At the faint end, TNG50 best agrees with observations in the near UV, in the optical u band and across the IR wavelengths. The largest disagreements, at the faint end, are in the central optical wavelength.
One case where the simulated LFs underpredict observations is the faint end of the SPIRE $\um{350}$ LF measured by \citet{Negrello2013}: however, as stated in Sec.~{\ref{subsubsec:obs_lf}} this LF is inferred in the very local Universe where the contamination and inhomogeneities can be large \citep{Marchetti2016}, whereas our sample more closely connects to the deeper observations from \citet{Marchetti2016}.

\begin{figure}
	\includegraphics[width=\columnwidth,keepaspectratio]{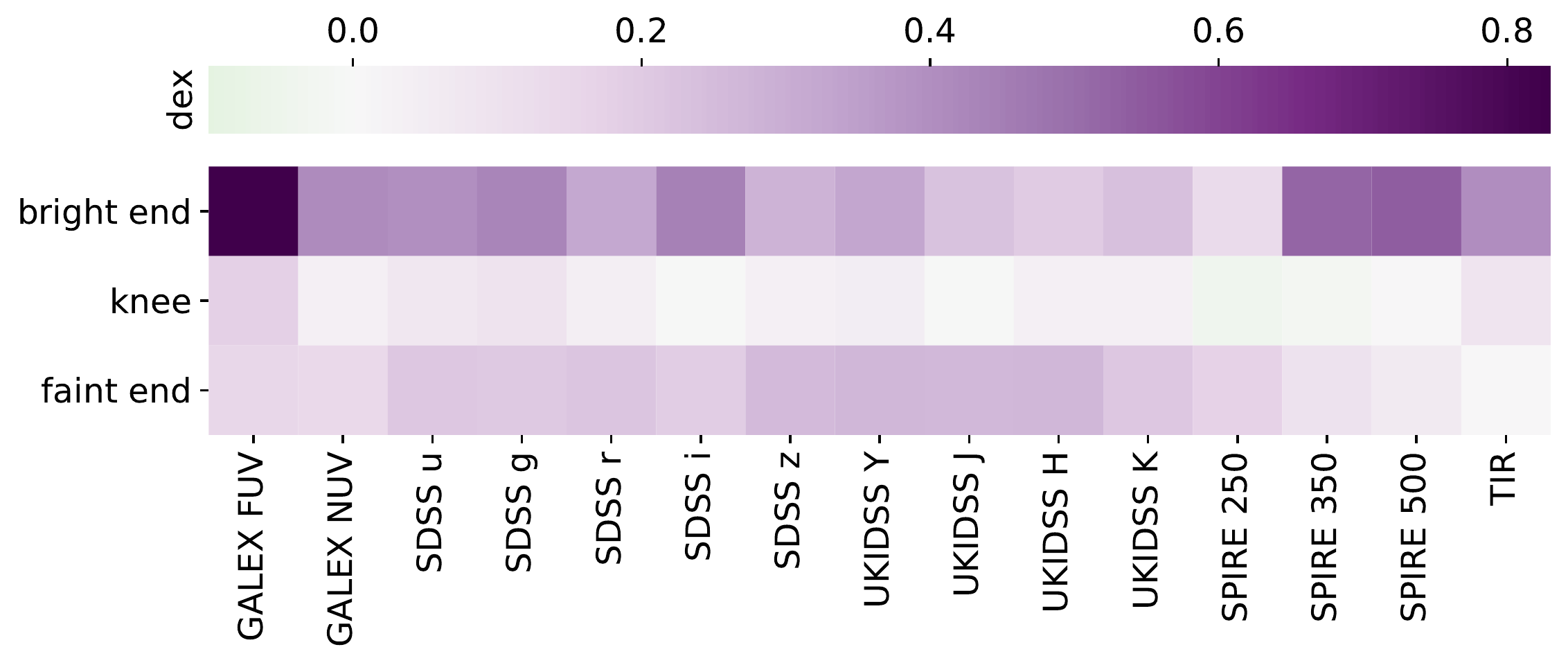}
     \caption{Median difference between the LFs of the TNG50 galaxies (with $5\,R_{1/2}$ apertures) and the observed LFs, at different wavelengths and at three luminosity regimes: $\pm$ 0.2 dex around the knee of the LFs and the brighter and fainter ends. The differences are in dex.}
    \label{fig:LF_diff}
\end{figure}

We quantify the level of (dis)agreement between TNG50 and observed LFs in Fig.~{\ref{fig:LF_diff}}. 
This figure shows the median discrepancy between observations in three luminosity regimes: $\pm$ 0.2 dex around the knee of the LFs in the brighter and fainter ends \citep[the position of the knee, acquired from][ is based on the characteristic luminosity of the Schechter function fit]{Driver2012, Marchetti2016}. 
Here the light from each galaxy is obtained by summing up all the contribution from within $5\,R_{1/2}$. 
Comparing the whole LFs, the FUV band shows the poorest agreement, mostly because of the disagreement at the bright end; the SPIRE 250 band instead shows the best agreement. 
In the UV and FIR domains, the TNG50 faint end has better agreement than the bright end, while in the optical and NIR only the knee of the LFs is reproduced to within 0.04 dex. 
The discrepancies do not seem to correlate with the wavelength, or only weakly. 
The shape of all LFs at the faint end closely follows that of the observations, where the upturn in some observational LFs at low luminosities \citep{Hill2010,Loveday2012,Driver2012} is visible in the simulations a s well. 
We will return to explore this feature in Sec. \ref{subsec:popul}, where we analyse different galaxy populations.

Since the TNG50+RT fiducial model shows discrepancies as high as 0.8 dex, in the rest of the paper, we explore different drivers of the galaxy LF shape and normalisation.

\section{Discussion}
\label{sec:disc}

\subsection{Discrepancy between the observed and TNG50 LFs}
\label{subsec:Discrepancy}

In order to disentangle the effects of different assumptions and settings of our approach, we individually investigate how aperture, orientation, RT recipe and simulation resolution affect the LFs. For the parameters that are not being discussed, we assume the fiducial values. For brevity reasons, we discuss only the most relevant options, however all of them are presented at \href{http://luminosity-functions.herokuapp.com/} {http://luminosity-functions.herokuapp.com/}.

\subsubsection{Aperture effect}
\label{subsubsec:LF-aper}

We have measured the TNG50 LFs corresponding to four distinct circular aperture choices in the plane of the sky: $5\,R_{1/2}$ (fiducial), $2\,R_{1/2}$, 30~kpc, and 10~kpc. The dust and stellar distributions are calculated inside $5\,R_{1/2}$ for all apertures in 3D. This means that, along the line of sight, we always collect light from a depth of $10\,R_{1/2}$, while the field of view is limited by the specific 2D aperture size (with an effective maximum being $10\,R_{1/2}$ as there are no sources beyond that point). 
Restricting the field of view without the limit on the line of sight, we mimic the observational apertures. The results in different bands are presented in Fig.~\ref{fig:LF_fidu}, in lighter colours compared to the fiducial model.

Comparing the aperture of 30 kpc to the fiducial one, as expected, the difference can be seen only at the bright end of the galaxy LF, which is now lower than when applying the fiducial aperture. A similar effect is observed for the $2R_{1/2}$\footnote{Not shown in the paper for clarity reasons, but see the link above.} and the smallest 10 kpc aperture. The discrepancy between the apertures is less pronounced moving towards the NIR, because the emission from the older stellar population \citep[that dominates the central galaxy region][]{Casasola2017}, has a higher contribution at these wavelengths. 

Several observational studies demonstrated that the bright end of the LF heavily depends on the aperture adopted in the magnitude calculation \citep{Hill2011, Bernardi2013}. This effect can be seen analysing the observational results from \citet{Loveday2012} and \citet{Driver2012} in the optical, where former applied Petrosian \citep{Petrosian1976} and latter Kron \citep{Kron1980} magnitudes on the same data set. While there is an offset between the two magnitude systems, they both can greatly underestimate galaxy fluxes \citep{Graham2005}. Additionally, in the UV domain, the observational data is derived inside the Kron aperture, while in the FIR the apertures are difficult to handle due to the poor resolution and high confusion noise \citep{Nguyen2010}.
In Appendix \ref{ape:petro} we test how our different aperture definitions correlate with the Petrosian aperture. 
The $5R_{1/2}$ aperture shows a high correlation, while for massive galaxies 30 kpc can be a good approximation. 
With the aperture of 10 kpc is expected to be too small to collect the the same amount of light as the Petrosian aperture.

We conclude that, while (only) the bright end of the galaxy LF is sensitive to the aperture choice, applying another relevant aperture definition (30 kpc), does not improve the results.

\subsubsection{Orientation effect}
\label{subsubsec:LF-ori}

Our data release contains, for every galaxy, fluxes for three galaxy orientations: edge-on, face-on, and random (given by the simulation). This allows us to also consider edge-on and face-on LFs. The random view is useful for the general comparison with observational surveys, while the analysis of specific orientations can aid in defining potential biases in observational studies \citep{Lovell2021, Yuan2021}.

\begin{figure*}
	\includegraphics[width=2\columnwidth]{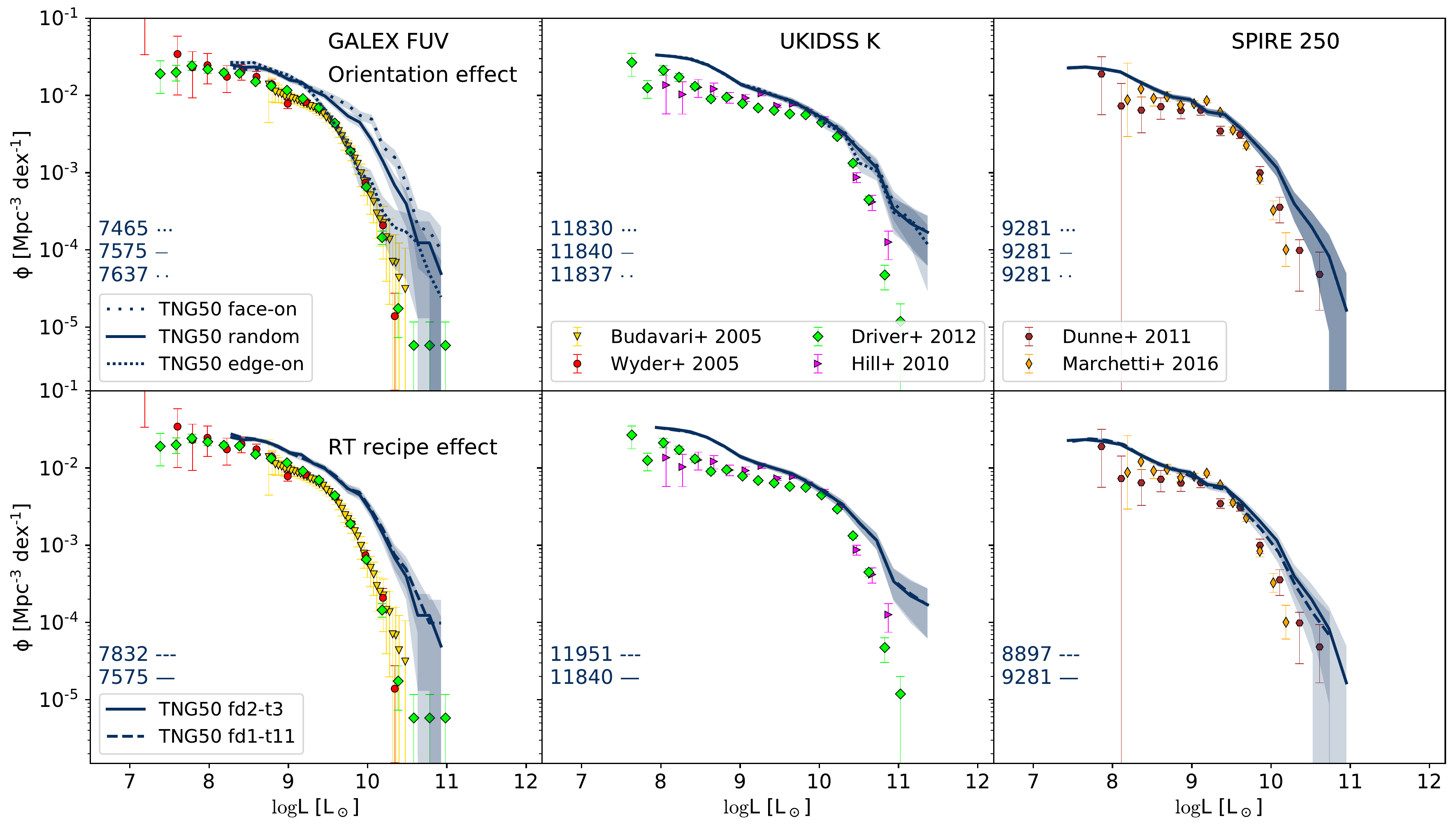}
     \caption{Similar as Fig.~\ref{fig:LF_fidu} but only for three bands and for the different galaxy orientations (RT recipes), top (bottom). 
     The solid line represents the fiducial model, while (loosely) dotted denotes (face-on) edge-on view, and dashed the fd1-t11 RT recipe.}
    \label{fig:LF_ori_rec}
\end{figure*}

The variability of the LFs with the galaxy orientation depends on the wavelength. Moving from UV towards IR, the effect declines because the stellar emission fades, and the dust attenuation is low at longer wavelengths. The top row of Fig.~{\ref{fig:LF_ori_rec}} shows the LFs in a selection of bands across the SED. In the UV, where the effect is most pronounced, we can see the impact of dust on the young stellar emission, as there is a considerable drop in the number of UV luminous objects as the inclination angle increases. These galaxies are increasingly star-forming, dusty, and largely asymmetrical. The effect is the same for every galaxy aperture.

If our sample included high-redshift and/or ultraluminous infrared galaxies (ULIRGs; $10^{12}L_\odot<L_{\mathrm{IR}}<10^{13}L_\odot$), that are compact, extremely star-forming and with hot dust \citep{Casey2014}
it would be necessary to incorporate the dust self-absorption in our RT procedure \citep{Camps2018, Ma2019}. For these objects, if they are disk-like, the results would then vary with the inclination in a wider wavelength range \citep{Lovell2021}.

In conclusion, since our observational samples do not favour any particular orientation, the fiducial model with the random view is the most compatible for the comparison. 
Furthermore, the effect of the orientation change can be only seen at low wavelengths.

\subsubsection{RT recipe effect}
\label{subsubsec:LF-rec}

One possible explanation for the discrepancies between observations and simulations could be that our RT settings are ill-chosen. Repeating the RT procedure on the whole TNG50 sample with different choices for the RT post-processing recipe is a highly resource-intensive job. However, to understand to what extent the recipe affects the LFs, we decided to repeat the procedure and recalculate the LFs for the {\tt{fd1-t11}} recipe. As stated in Sec. \ref{subsec:cal}, the fiducial {\tt{fd2-t3}} recipe and the {\tt{fd1-t11}} recipe differ in both the diffuse and compact dust: the former has larger amounts of diffuse dust, while the latter assumes more screened H\textsc{ii} regions.

The bottom panel of Fig.~{\ref{fig:LF_ori_rec}} shows that the choice of the RT recipe has a very modest effect on the LFs. The recipe with low $f_{\text{dust}}$ displays an emission excess across the UV and optical range, which demonstrates that, in our setting, the diffuse dust is driving the results.
In the FUV the recipes agree with each other despite the difference in the $f_{\text{dust}}$ and $\tau$.
In this band the emission predominantly comes from young stars, therefore the additional coating of the \textsc{hii} regions can compensate for the lack of diffuse dust. 
Consequently, the {\tt{fd1-t11}} recipe has slightly lower IR emission than the fiducial recipe. Overall, however, the change of RT recipe does not bring the simulated LFs more in agreement with the observational data.
The same conclusions are derived if we compare the RT recipes assuming different aperture definitions.

\subsubsection{Hydrodynamic simulation resolution effects}
\label{subsubsec:LF-sim}

The TNG simulations are realised with and without baryons, at a number of resolutions and box sizes \citep[for an overview, see Table 1 in ][]{Nelson2019a}. 

The TNG model was designed at a resolution similar to the one of the original Illustris simulation, i.e, at the resolution of the other flagship run of the IllustrisTNG project, TNG100 \citep{Pillepich2018b,Nelson2018,Naiman2018,Marinacci2018,Springel2018}. 
The TNG100 and TNG50-2 simulations adopt comparable resolutions.
The other resolution simulations were run by assuming exactly the same physical model and subgrid parameter values\footnote{This is the case but for a) the choices of the softening lengths and of the black hole kernel-weighted neighbour number \citep{Pillepich2018a} and b) a somewhat steeper dependence of the star formation rate on gas density only for the densest gas, in TNG50 \citep{Nelson2019b}.}  \citep{Pillepich2018a}. 
This approach can manifest as quantitative changes in the simulations’ outcome for different adopted numerical resolutions, to different degrees depending on the galaxy properties under scrutiny \citep{Pillepich2018a,Pillepich2018b,Pillepich2019}.
To explore how the resolution of the cosmological hydrodynamical galaxy simulation influences our results, we have repeated our analysis for the TNG50-2 simulation, which has 8 (2) times worse mass (spatial) resolution than the flagship TNG50 run. 
For this, we have assumed and applied the same fiducial RT modeling (i.e. adopted  the same calibrated RT parameters as for TNG50) and recalculated LFs starting from the data simulated by TNG50-2. 
As it has been shown that, within the TNG model, an improved resolution  may result in somewhat larger galaxy masses and SFRs \citep{Pillepich2018a,Pillepich2018b, Pillepich2019, Donnari2019}, we can expect to see a downshift in the LFs when moving from TNG50 to TNG50-2.

\begin{figure*}
	\includegraphics[width=1.97\columnwidth,keepaspectratio]{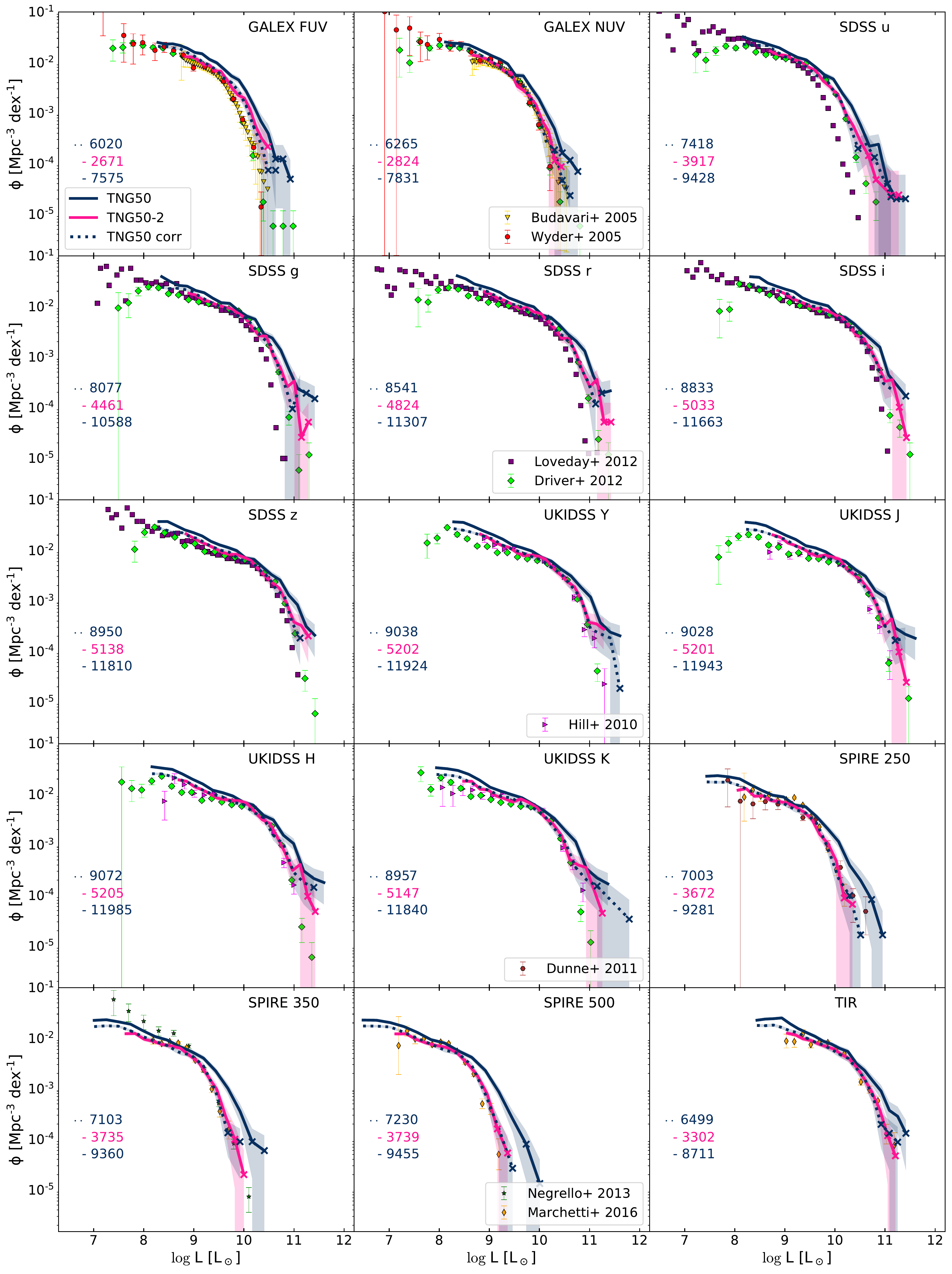}
     \caption{Same as Fig. \ref{fig:LF_fidu} but comparing the results from TNG50 (navy blue) to a worse-resolution counterpart called TNG50-2 (pink), with 8 (2) times worse mass (spatial) numerical resolution, both post-processed with the same fiducial RT modeling. The navy blue dotted line represents a possible way to correct the TNG50 results to account for the effects of resolution. In all curves, light from each galaxy is taken from a $5\,R_{1/2}$ aperture. We give the LFs from the simulated galaxies only above the completeness limit imposed by the selection: stellar mass $>10^{8}\; \mathrm{M_\odot}$ ($10^{8.6}\; \mathrm{M_\odot}$) for TNG50 (TNG50-2).}
    \label{fig:LFs_resoN}
\end{figure*}

\begin{figure}
	\includegraphics[width=\columnwidth,keepaspectratio]{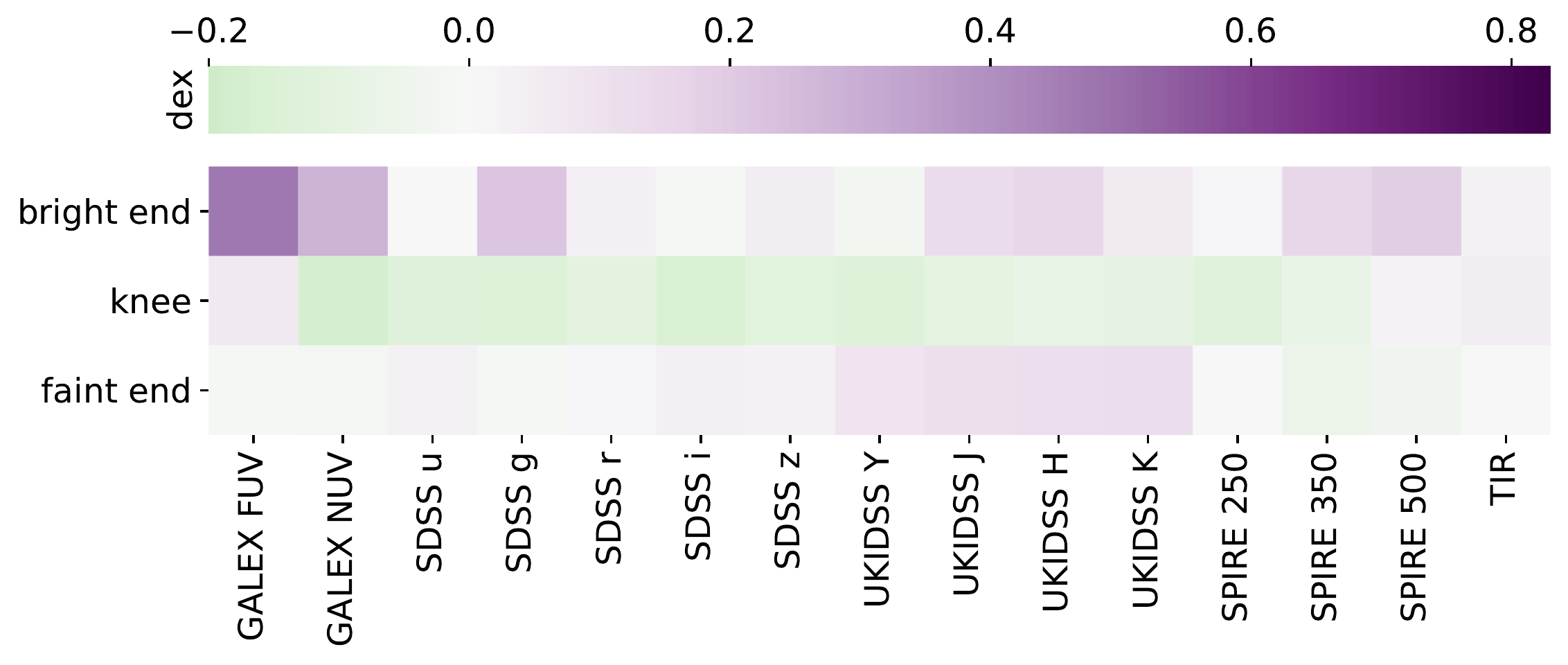}
     \caption{Same as Fig.~\ref{fig:LF_diff} but for the TNG50-2. The colour bars are showing the same range as in Fig.~\ref{fig:LF_diff}.}
    \label{fig:LF_diff2}
\end{figure}

The results are presented in Fig.~{\ref{fig:LFs_resoN}}. As expected, we see that TNG50-2 has lower LF values across the entire wavelength range. The agreement with the observed LFs is substantially improved, which we quantify in Fig.~{\ref{fig:LF_diff2}}. This finding reveals that the two simulation runs behave in the same way in the observed space and across the whole wavelength range as noticed in the physical space. Therefore, we expect the same high level of agreement of the galaxy LF between TNG100 and observations, since the stellar masses of halos, and so also the stellar mass function, decrease with worse numerical resolution (as discussed e.g. in \citet{Pillepich2018a} Appendix A on the TNG model convergence, and \citet{Pillepich2018b} Appendix A on the TNG100/300 convergence).

From the comparison between TNG50 and TNG50-2, we can derive a resolution correction to be applied to TNG50. Following \citet{Pillepich2018b,Engler2021} and \citet{Vogelsberger2018, Vogelsberger2020b}, and assuming that the resolution change does not affect the halo mass, we first subdivide the sample based on the halo mass. In each halo mass bin, we calculate the total luminosity in each band and redshift and for both simulation runs.
The ratio of these values is the factor we use to scale the luminosity of each individual galaxy. The outcome of this exercise is shown as the navy blue dashed lines in Fig.~{\ref{fig:LFs_resoN}}, where we can see that the transition from TNG50 to TNG50-2 is excellent. 
An alternative way to improve the agreement with observations could have been obtained by adopting different choices in the TNG model (i.e. different subgrid parameter values) to account for the resolution change, prior to run the galaxy simulation, as done e.g. within the EAGLE project \citep{Schaye2015, Crain2015}  and demonstrated by \citet{Mitchell2021}.

In the final overview of the simulated (both TNG50-rescaled and TNG50-2) and observed LFs, in most bands, we see an outstanding agreement. The minor excess at the very bright end is partially caused by the low statistics in the high luminosity bin. 
However, in the FUV band there is still an overabundance of luminous galaxies. 
This can still be attributed to the uncertainty about where i.e. with which aperture the observations are taken.
At the same time, from Fig.~{\ref{fig:seds_cig_ssfr}} it can be seen that for the majority of recipes, the flux in the FUV band is typically much higher than the one in the NUV band, while the DustPedia sample has the opposite trend in most panels. In their previous work \cite{Baes2019} and \cite{Trcka2020} analysed the mock fluxes of the EAGLE simulations, derived in a similar manner as we employ here. These studies revealed that the UV part of the spectrum exhibits the highest discrepancy compared to the observations also in the case of EAGLE.
This wavelength regime is difficult to constrain because, while the emission in the FUV emerges from the star-forming regions, they are still not resolved in the cosmological simulations. Therefore, the use of templates, like the \textsc{mappings-iii} templates we use here, is necessary. 
While we improved the method by introducing more realistic settings (time-dependent covering factor, variable compactness), we argue that there is a need for templates that can more successfully exploit the data from the simulations (Kapoor et al., in prep). 

\subsection{Contribution of different galaxy populations to the LF}
\label{subsec:popul}

We conclude the paper by analysing how distinct galaxy populations contribute to the LFs. We split our sample according to three different criteria: morphology, environment and star formation activity. 

\begin{figure*}
	\includegraphics[width=1.9\columnwidth]{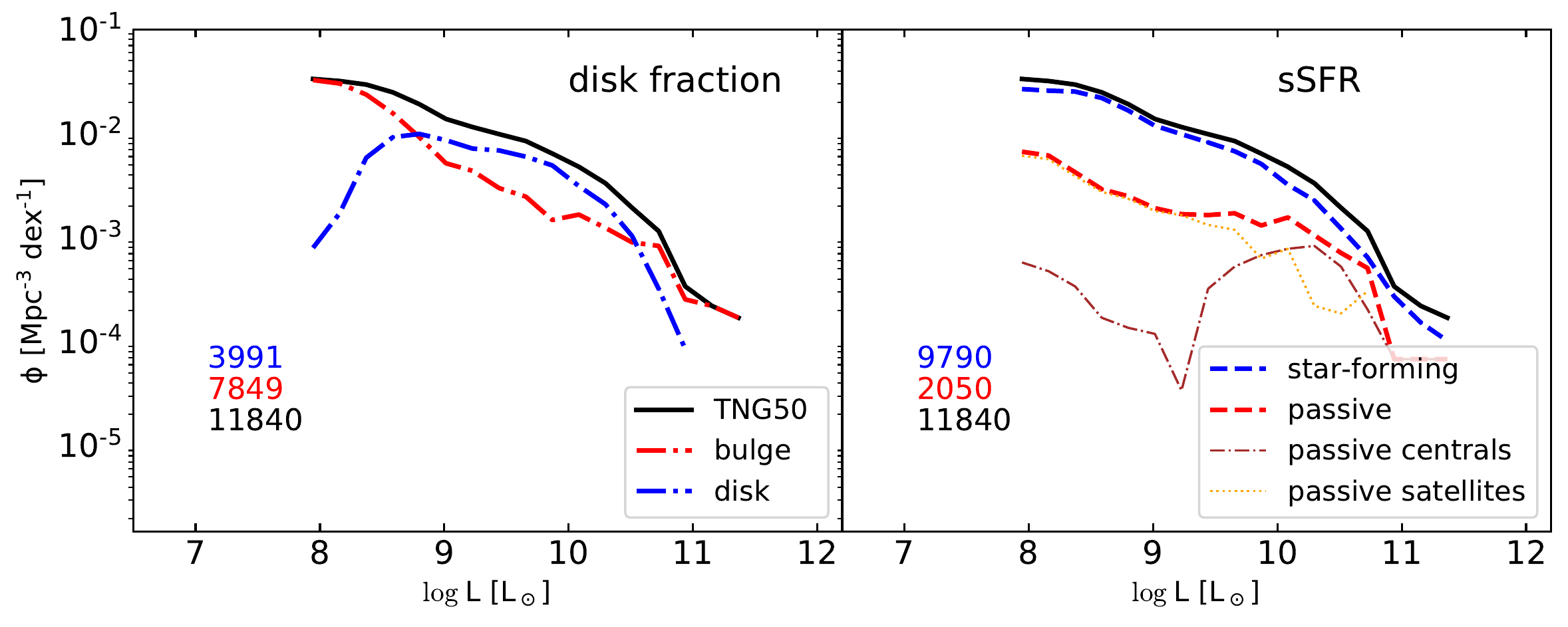}
     \caption{TNG50 LF in the UKIDSS K band at $z\le 0.1$. The black line in both panels is the TNG50 LF inside $5\; R_{1/2}$ for the random orientation and fiducial RT recipe.  
     \textit{Left:} The red and blue dash-dotted lines correspond to the galaxies defined as a bulge and disk-dominated, respectively.
     \textit{Right:} The red and blue dashed lines show how the sample splits according to the sSFR. The brown dot-dashed and orange dotted lines show the passive centrals and the passive satellites, respectively.
     }
    \label{fig:discuss}
\end{figure*}

For the morphology, we apply a simple prescription to divide the sample into disk and bulge-dominated galaxies, by exploiting the stellar kinematical data \citep{Genel2015}. Based on their angular momentum, stars are assigned to either disk or bulge, and for a galaxy to be defined as disk-dominated, the mass fraction of disk stars should be over 0.35 \citep{Hoffmann2020}. We present the results in the left panel of Fig.~{\ref{fig:discuss}}, where we show the luminosity function in the UKIDSS K band. Since this band traces the emission of evolved stars and is insensitive to the SFR, this LF is a proxy for the stellar mass function. Bulge-dominated galaxies are prominent in both the bright and faint end of the LF, while disk galaxies dominate the LF at intermediate luminosities. \citet{Trayford2019} derived comparable results for the stellar mass function of the EAGLE simulations. Comparing to observations, \citet{Kelvin2014} calculated LFs for the GAMA survey and noticed a similar shift in the contribution of the different Hubble populations at the luminous end, although their sample contained mostly disk-like galaxies, contrary to our results. When we repeat the same exercise for the TNG50-2 simulation, we find that the number of disk galaxies decreases even more, as it is shown that the low resolution simulations have thicker disks \citep{Pillepich2019}.
Analysing the visual morphologies of the TNG100 galaxies via machine learning techniques, \citet{Huertas2019}  derive similar results for the mass function, however their population split shows a lack of early type galaxies with high masses.

Moving to the star formation activity, we divide the sample into star-forming and passive galaxies, applying a cut in sSFR. Following \citet{Salim2014}, we identify galaxies as star-forming if ${\text{sSFR}} > 10^{-10.8}$~yr$^{-1}$, and passive otherwise. We present the results in the right panel of Fig.~{\ref{fig:discuss}}. 
We see that the upturn, seen in the global LF at the faint end, is mainly driven by the shape of the LF for the passive galaxies. The existence and the origin of the upturn towards the faint end of LFs (and mass functions) have been heavily debated \citep[e.g.][]{Blanton2003, Hill2010, Loveday2012, Driver2012}. \citet{Taylor2015} review different ways of subdividing a galaxy sample into blue and red galaxies, depending on the global colours. 
They argue that the reason a few studies show the presence of this upturn in the population of red galaxies is the contamination by the blue galaxies, due to inadequate sample splitting. We tested this hypothesis by changing our threshold in the sSFR and employing three different colour and stellar mass cuts \citep{Bell2003, Baldry2004, Peng2010}, but the shape of the passive LF remains the same. This result is not unexpected if we look at Fig.~{\ref{fig:select}}, where there is a clear bimodality in the stellar mass for the passive galaxies. 
To inspect this further, we split the passive sample into centrals and satellites\footnote{A galaxy is defined as central if it resides in the lowest potential within a dark matter halo. It is typically the most massive galaxy in its halo.}, which revealed that most of the emission at the luminous end comes from the centrals, while the upturn is completely dominated by the passive satellites, in agreement with observations \citep{Lan2016}.
The number of passive centrals, compared to the number of the passive satellites at low luminosities (stellar masses) is insignificant. 
This is a consequence of a generally small ratio of passive central galaxies at low masses \citep{Geha2012}, that is captured by the TNG50 simulations.
As shown in, e.g. \citet{Donnari2021} and \citet{Joshi2021}, going to worse resolutions (e.g. TNG50-2) will show an increased fraction of quenched low mass central galaxies.

Our findings demonstrate that different galaxy populations affect in a complex manner the LFs in simulations, in accordance with the observations.

\section{Summary and Conclusions}
\label{sec:conc}

In this study, we have produced mock fluxes over a wavelength range from FUV to sub-mm for low-redshift galaxies from TNG50, a state-of-the-art cosmological hydrodynamical simulation for galaxy formation. Using the \textsc{skirt} code we have calibrated the RT procedure on a small TNG50 subsample that corresponds to the observational DustPedia sample. The calibration constrains the amount of diffuse dust and the dust covering of the SF regions. The calibration was based on the comparison of various luminosity scaling relations and full galaxy SEDs (Figs.~\ref{fig:LLLLLcig} and \ref{fig:seds_cig_ssfr}). Both the simulated subsample and the observed data from the DustPedia sample are processed using the \textsc{cigale} SED fitting code to ensure the comparison is consistent.

We applied the calibrated RT procedure to the whole TNG50 sample (and the worse-resolution TNG50-2 sample) and produced the observed fluxes and rest frame magnitudes in 53 broadbands for $z\le 0.1$, for three orientations and four apertures. Full galaxy SEDs are also generated (Table~\ref{tab:1}). These data are publicly released (\href{https://www.tng-project.org/trcka22}{https://www.tng-project.org/trcka22}) and we invite interested colleagues to use these data for their own science projects.

We have constructed and analysed the LFs for the TNG50 simulation over a range of broadband filters covering UV to submm wavelengths and compared them to different observational data. For our fiducial setting, we found agreement on the overall shape of the LFs and the normalization at the knee, however, the simulations mainly overestimate the observations (Fig.~\ref{fig:LF_fidu}), with the best agreement for total IR at faint end ($\sim$ 0.004 dex) and with the strongest discrepancies in the FUV ($\sim$ 0.8 dex) at the bright end.

We varied several parameters in order to investigate their effect on the LFs, and we derived the following conclusions:
\begin{itemize}
    \item Our results are sensitive to the aperture choice at the LF bright end across the wavelength range. 
    However, the choice of the aperture seems not to be the cause of the discrepancy between simulated and observed data at the faint end (Fig.~{\ref{fig:LF_fidu}}), and not even at the luminous end, unless observations are de facto sensitive to effective radial apertures as small as 10 kpc or so. 
    Therefore, it is difficult to resolve the discrepancy between simulated and observed data via aperture effects alone. 
    \item By changing the orientation of galaxies we find that the effect is wavelength dependent.
    The strongest differences in the LFs between the face-on and the edge-on views can be seen in the UV bands (Fig.~\ref{fig:LF_ori_rec}) whereas inclination effects are negligible at other wavelengths. 
    \item Changing the recipe of our RT procedure to incorporate less diffuse dust and more covered SF regions only marginally affects the outcome. In most bands, the diffuse dust is the principal driver of the results (Fig.~\ref{fig:LF_ori_rec}).
    \item  Applying the same RT fiducial modeling to the TNG50-2 run — with a worse numerical resolution than TNG50 i.e. similar to the other flagship run TNG100 — results in a better agreement with the observations. This is a manifestation of the fact that the TNG model has been designed and chosen at approximately the TNG50-2 or TNG100 resolution, with no changes of model parameters at different resolutions: the TNG model hence appears not fully converged at the TNG50 level (Fig.~\ref{fig:LFs_resoN}). Based on the comparison between the TNG50 and TNG50-2 outcomes, it is possible to devise a resolution correction on the luminosities of the high-resolution TNG50 galaxies that brings their LFs in excellent agreement with the observed ones.

    \item The final simulated LFs (corresponding to TNG50-2 simulation or the rescaled TNG50 simulation) show a high level of agreement across the wavelength range.
    \item We dissect the UKIDSS {\em{K}} band LF (tracer of the stellar mass function), and discuss the effect of the (kinematically driven) galaxy morphology and star formation. The noticed effects of different galaxy populations correlates favourably with previous studies of both observations and simulations (Fig.~{\ref{fig:discuss}}):
    \begin{itemize}
        \item Splitting the galaxy sample into bulge and disk dominated reveals that disk galaxies mostly populate moderate luminosities, while the elliptical galaxies are residing at the extreme ends of the galaxy LF.
        \item Analysing separately low and high sSFR galaxies, we see that, while the dominant population is the star-forming one, the upturn in the galaxy LF is driven by the passive component.
    \end{itemize}
\end{itemize}

The RT procedure applied in this work, while improving, still has not converged to its optimal version. To help reduce the current discrepancies in the predicted FUV/UV fluxes, one could consider developing more realistic RT subgrid models for the simulated SF regions \citep[potentially based on the high resolution hydro dynamical simulations of SF regions][]{Kim2017,Duarte2017,Seifried2018,Smith2020,Kannan2020,Kannan2021,Smith2021,Tacchella2022}. These models should more accurately represent the complex, clumpy structure and physical properties of the gas and dust immediately surrounding the young stellar objects. On another note, in the interest of extending the analysis to higher redshifts, the potential evolution of the dust-to-metal ratio has to be accounted for \citep{Peroux2020,Vogelsberger2020b,Popping2022}.

While future cosmological galaxy simulations will generate highly resolved galaxies in even greater numbers, the ideal option to test and learn from them demands comparing them with observations. In this study, we applied the RT procedure in the post-processing, however, we suggest that the incorporation of the RT procedure directly in the calibration or design of the cosmological hydrodynamical galaxy simulations would be greatly beneficial. It would provide a consistent comparison with the observations, and aid constraining of the subgrid recipes.

\section*{Acknowledgements}
DN acknowledges funding from the Deutsche Forschungsgemeinschaft (DFG) through an Emmy Noether Research Group (grant number NE 2441/1-1).
The primary TNG simulations including TNG50 were carried out with compute time granted by the Gauss Centre for Supercomputing (GCS) under Large-Scale Projects GCS-ILLU and GCS-DWAR on the GCS share of the supercomputer Hazel Hen at the High Performance Computing Center Stuttgart (HLRS).
The computational resources (Stevin Supercomputer Infrastructure) and services used in this work were provided by the VSC (Flemish Supercomputer Center), funded by Ghent University, FWO and the Flemish Government -– department EWI
This research made extensive use of the NumPy, MatPlotLib, Plotly and Pandas Python packages.

\section*{Data Availability}

The DustPedia data used in this study can be acquired via \href{http://dustpedia.astro.noa.gr/} {http://dustpedia.astro.noa.gr/}. The simulation data used to derive fluxes is retrieved from \href{https://www.tng-project.org/data/}{https://www.tng-project.org/data/}. 
The complete results of this paper, including full galaxy SEDs and the broadband fluxes at $z\leq 0.1$ are made available at \href{https://www.tng-project.org/trcka22}{https://www.tng-project.org/trcka22}.
The luminosity function plots are  presented on \href{https://luminosity-functions.herokuapp.com/}{https://luminosity-functions.herokuapp.com/}



\bibliographystyle{mnras}
\bibliography{bib} 




\appendix
\section{Aperture effects}
\label{sec:A-aper}

During our sample selection phase we employed the aperture of $2\, R_{1/2}$ while we used $5\,R_{1/2}$ for our further analysis.
We settled on this arrangement because the former aperture was already available from the subhalo catalogue, thus we avoided the expensive treating of the galaxy particle data in this phase of the project.
However, we inspect to what extent the stellar mass and the SFR vary with this aperture change.

\begin{figure}
	\includegraphics[width=\columnwidth]{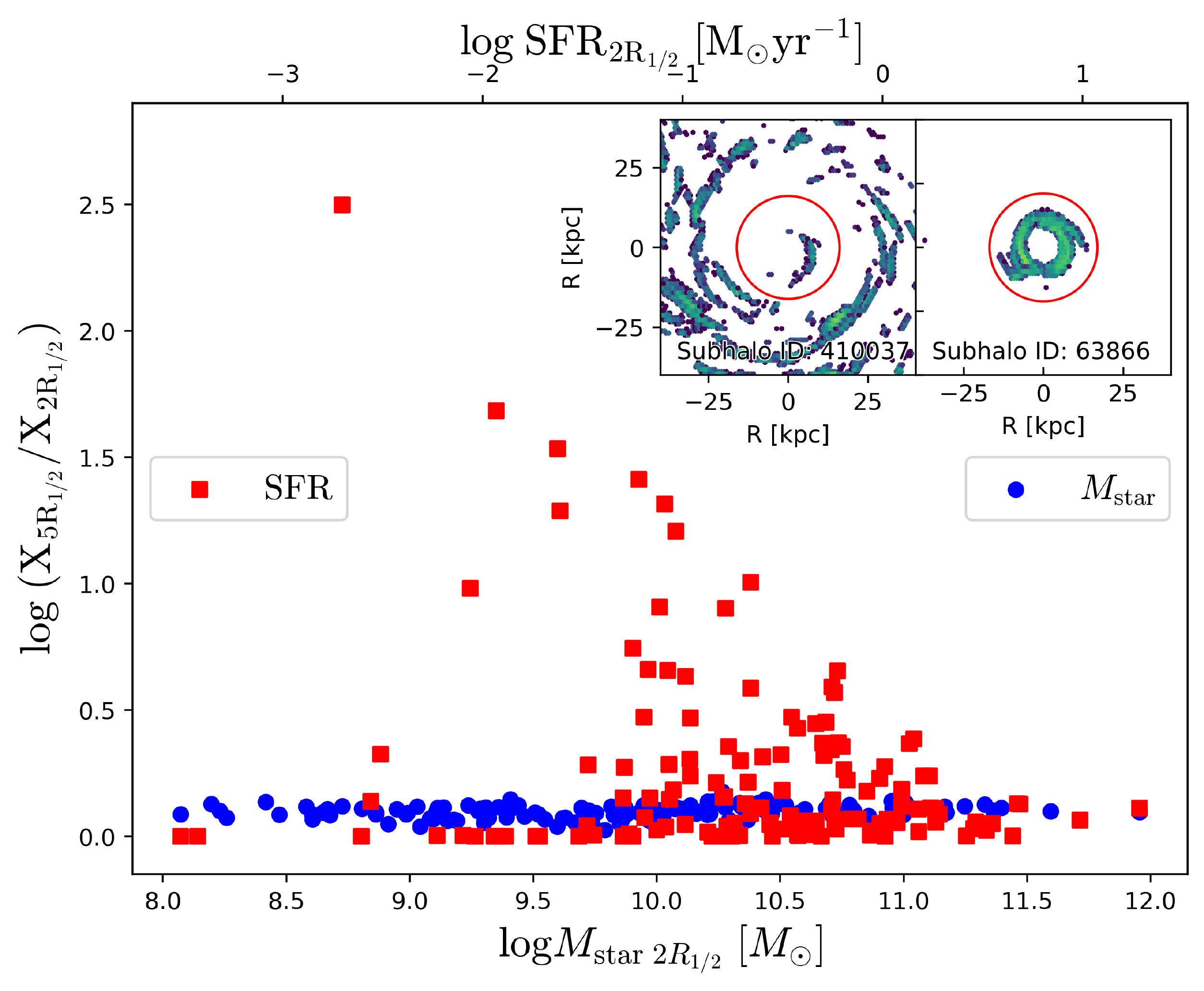}
     \caption{Ratio of the stellar mass (blue) and SFR (red) inside two different apertures. $70\%$ of the sample has the difference in SFR lower than 0.25 dex. \textit{Inset}: Density of the SF gas for two galaxies with $M_{\mathrm{star}}\approx 10^{11}\; \mathrm{M_\odot}$ and $2R_{1/2}\approx 16.5$ kpc (red circle). The mass of the gas outside the circle contributes to the discrepancy in SFR.}
    \label{fig:aper}
\end{figure}

The results are shown in Fig.~\ref{fig:aper}. As expected from the aperture definition, the increase of the stellar mass with the aperture is limited - always less than 0.2 dex and 0.1 on average. 
However, gas and dust can reside at much larger radii due to the feedback outflows, therefore, the SFR for some galaxies changed drastically (although 20 per cent of galaxies has an increase in SFR less than 0.01 dex). 
The example of this can easily be seen in the inset of the same figure, where we show two galaxies with similar stellar mass and stellar half mass radius, but different distribution of the star-forming gas. 
The galaxy on the right has almost all star-forming gas inside $2R_{1/2}$ (red circle), therefore the difference between the SFRs will be minimal. 
On the contrary, the one on the left has most of the gas outside $2R_{1/2}$.
Analysing how this discrepancy affects the distributions of stellar mass and sSFR, we found that the medians shift towards higher values by 0.07 dex for both properties.
Since these global distributions do not deviate greatly, we are confident that this aperture change will not affect the RT calibration.

\section{RT procedure statistics and model variations}
\label{sec:A-tests}
In this section, we investigate the quality of our RT procedure as well as the differences in the SEDs following the change of a number of parameters. All plots are derived assuming the fiducial recipe: $\tau =3$ and $f_{\mathrm{dust}}=0.2$ for all galaxies in the TNG50 subsample.

\begin{figure*}
	\includegraphics[width=1.8\columnwidth]{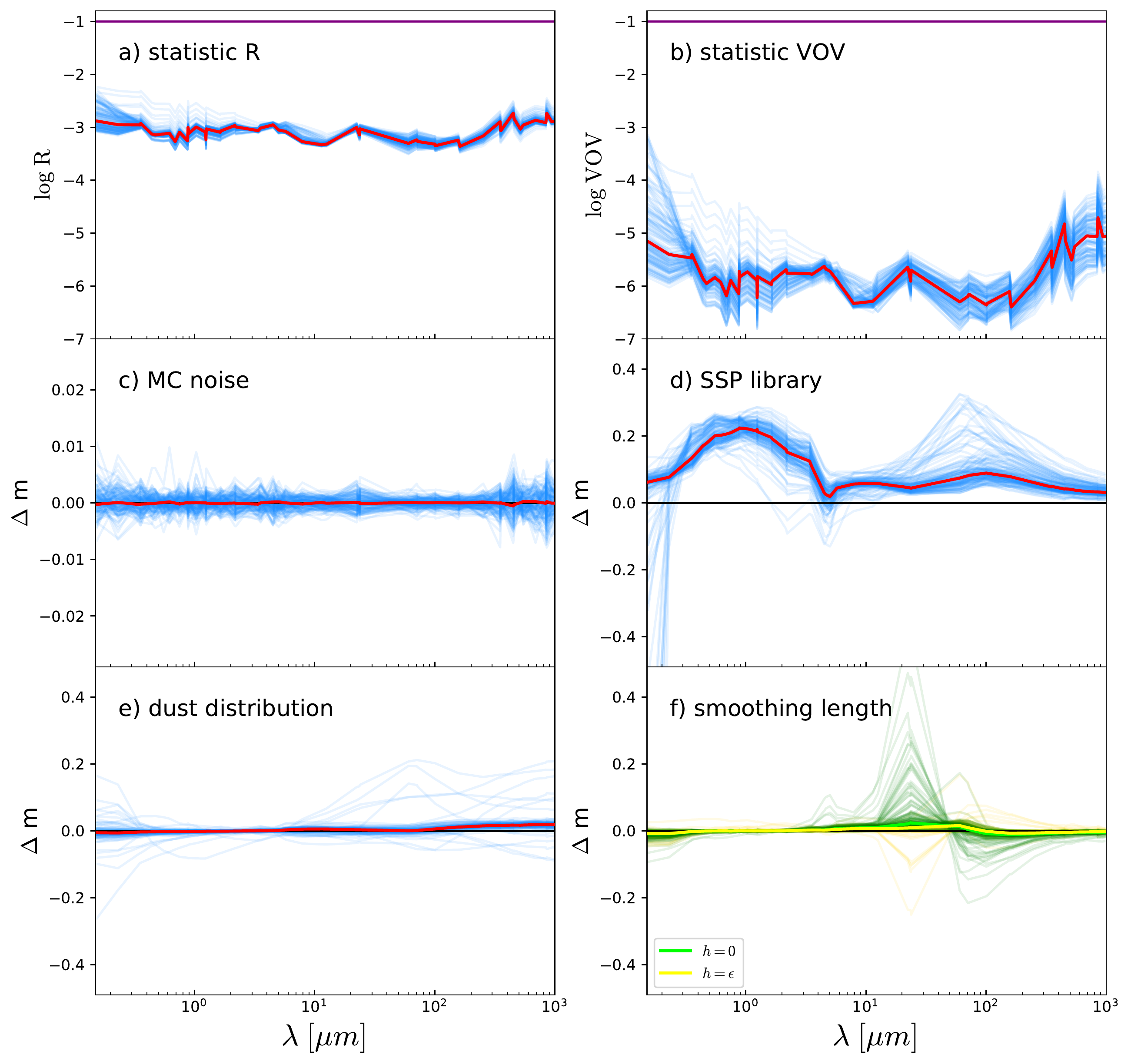}
     \caption{
     \textit{a}) and \textit{b}): 
     Results of the statistical tests, the relative error R, and the variance of the variance VOV. 
     \textit{c}):
     Monte Carlo noise shown as the magnitude differences of the different bands for the two identical RT simulation runs.
     \textit{d}), \textit{e}) and \textit{f}):
     Change in galaxy magnitudes due to the different SSP library, dust distribution and smoothing length, respectively.
     The blue (green/yellow for \textit{f}) lines represent individual galaxies, and the red line is the median of the TNG50 sumbsample.
     The purple line in the top panels shows the reliability threshold.
     }
     
    \label{fig:all_tests}
\end{figure*}

We performed two statistical tests (already implemented in \textsc{skirt} \citep{Camps2020}), and obtained the relative error, R and the variance of the variance, VOV. 
The test results are shown in Fig.~\ref{fig:all_tests} a) and b).
Both statistics in each point are derived from the high order sums of the photon packet contributions \citep{Camps2020}. 
As the recommended maximum value for both R and VOV is 0.1, our results are reliable in the whole wavelength range.
Additionally, we run the supplementary test to see the scale of the Monte Carlo noise by running the simulations with the same parameters twice. 
The results are shown in Fig.~\ref{fig:all_tests} c), where we see that the difference in the broadband fluxes is always below 0.02 mag.  

In our further tests, we compare the resulting SEDs derived from Starburst99 SSP \citep{Leitherer1999} with IMF from \citet{Kroupa2001}, and the SSP from \cite{Bruzual2003} with \cite{Chabrier2003} IMF. 
These are shown in Fig. \ref{fig:all_tests} d). 
As expected, the effect is wavelength dependent with the highest median deviation in the optical (0.23 mag). 
Since Starburst99 is specifically designed for the treatment of massive stars \citep{Leitherer1999}, they are more appropriate to use exclusively for the star-forming regions.
Furthermore, given that the SF regions are dominated by massive stars, the difference between \citet{Kroupa2001} and \cite{Chabrier2003} IMF is going to be limited.

As for the diffuse dust, we explore if inserting randomness in the $f_{\mathrm{dust}}$ of the distinct gas cells, with an unchanged total galaxy $M_{\mathrm{dust}}$, would affect a galaxy SED. 
In this test, we draw from the uniform distribution (from 0.01 to 0.4) to assign a value to each gas cell, and the results are shown in Fig.~\ref{fig:all_tests} e). 
The highest deviation of the median line is 0.022 mag.
The individual galaxies that have higher discrepancies are low dust mass and passive galaxies, where every change in the dust distribution will be evident.

Finally, for the argumentation about the smoothing length choice, we performed tests treating (1) all sources as point objects ($h_{\mathrm{sl}}=0$) and (2) with the smoothing length equal to the simulation softening length ($\epsilon$) \citep{Torrey2015},
We compared these with our fiducial smoothing length, corresponding to the radius of the sphere enclosing 32 stellar particles, in Fig. \ref{fig:all_tests} f).
The two alternatives evince a similar (somewhat stronger for $h_{\mathrm{sl}}=0$) effect with an increased UV attenuation and emission at 22 $\mu$m.
This is expected as it is more difficult for the now more compact stellar emission to escape, therefore it is heating dust to higher temperatures.
The maximum deviation of the sample median for the run with $h=0$ is 0.024 mag  and 0.016 mag for the run with $h=\epsilon$.

\section{Comparison of physical properties}
\label{sec:A-comp-phy}
To validate our post processing procedure, we confront the true galaxy properties with those derived from the \textsc{cigale} SED fitting tool and from the common tracers.
\subsection{\textsc{cigale} stellar mass and SFR}
\label{sec:A-tng-cig}

\begin{figure*}
	\includegraphics[width=1.8\columnwidth]{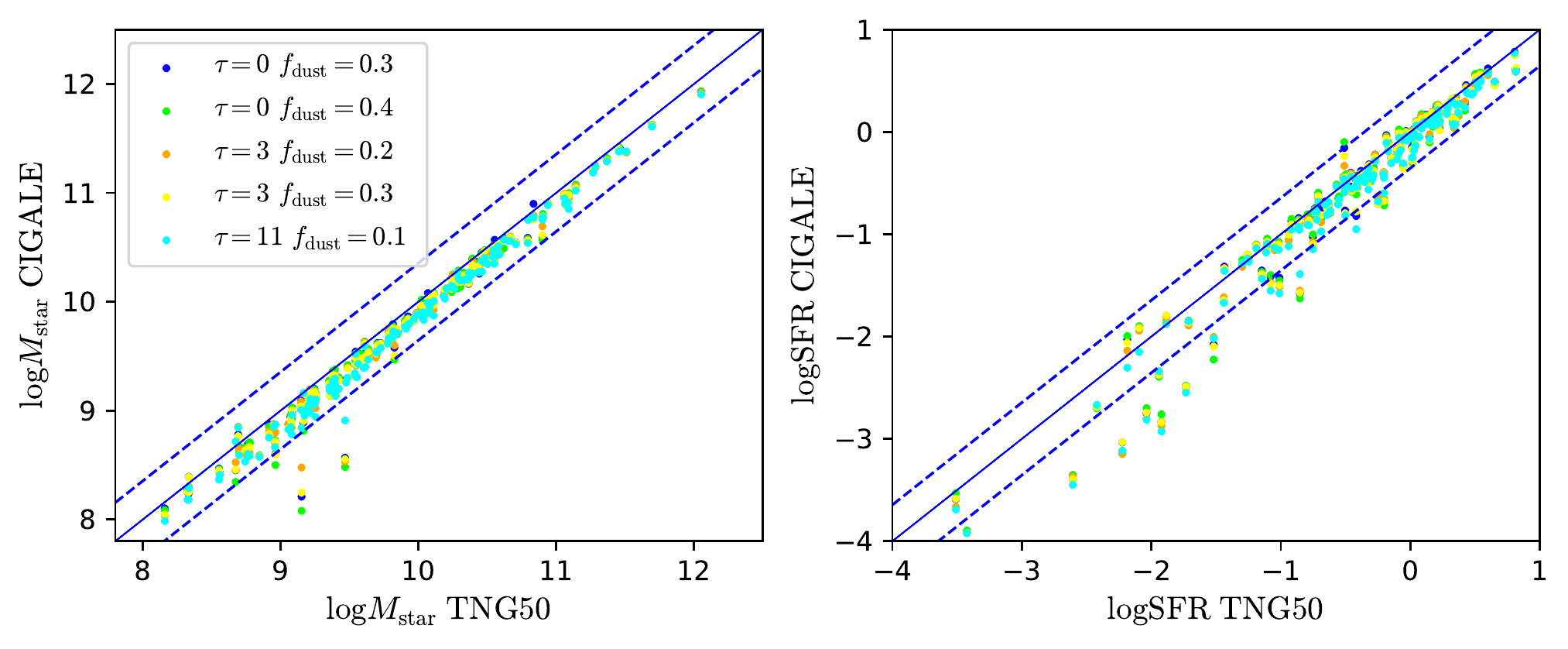}
     \caption{Comparison between $M_{\mathrm{star}}$ (left) and SFR (right) retrieved from the simulation data and from \textsc{cigale}, both inside $5\,R_{1/2}$, for different recipes. The blue solid line is one-to-one line, and the dashed lines represent $\pm 0.25 \; \mathrm{dex}$ lines. 
     }
    \label{fig:TNG_CIGALE}
\end{figure*}

In Fig. \ref{fig:TNG_CIGALE} we analyse galaxy stellar mass (left) and SFR (right) for 136 galaxies of the TNG subsample. 
Both SFRs represent instantaneous SFRs.
The different colours correspond to the different recipes considered in this study and mentioned in the main text.
Increasing $\tau$ generally increases the spread in the relations, as the young stellar emission is being dimmed.

\subsection{Stellar mass and SFR tracers}
\label{sec:A-tng-prox}
\begin{figure*}
	\includegraphics[width=1.8\columnwidth]{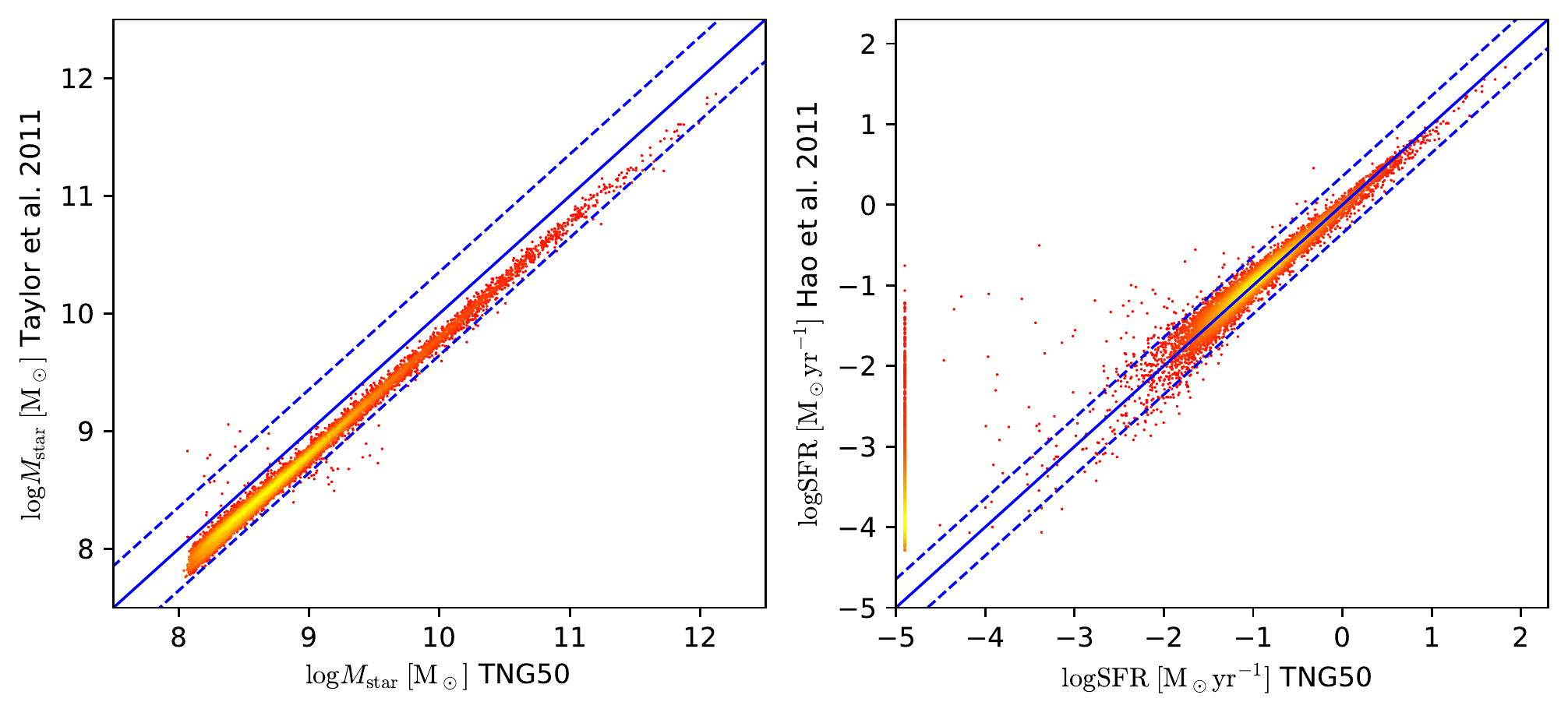}
     \caption{Comparison between $M_{\mathrm{star}}$ (left) and SFR (right) retrieved from the simulation data and from the fluxes, based on the recipe from the literature. 
     The total TNG50 sample at $z=0$ is presented. 
     The galaxies at the left corner of the SFR plot are those with intrinsically $\mathrm{SFR}=0$.
     The blue solid line is one-to-one line, and the dashed lines represent $\pm 0.25 \; \mathrm{dex}$ lines. 
     }
    \label{fig:TNG_prox}
\end{figure*}

Here we explore how the intrinsic simulation properties correlate with the appropriate proxies from the literature.
The results are shown in Fig. \ref{fig:TNG_prox}.
For the $M_{\mathrm{star}}$ tracer we rely on the recipe  from \citet{Taylor2011}, based on the optical measurement i.e. $\log( M_{\mathrm{star}}/M_{\mathrm{\odot}}) = 1.15+0.7(g-i)-0.4M_i$.
Other available recipes provide different scaling coefficients, and our results are within these uncertainties \citep{Gallazzi2009, Zibetti2009, Roediger2015}. 
As a SFR proxy, the FUV band is the most common choice. 
We selected a dust-corrected recipe from \citet{Hao2011}, which also includes MIPS 24 $\mu m $.
Although this proxy traces longer-term SFR \citep[$\sim$ 100 Myr][] {Kennicutt2012}, the agreement is outstanding.
To check the difference between instantaneous and  time-averaged SFR (over 100 Myr), we used the supplementary TNG50 catalogue \citep{Donnari2019,Pillepich2019}, and compared these SFRs within the aperture of $2R_{1/2}$.
We found that the instantaneous SFR deviates from the time-averaged only in the low SFR regime and no more than 0.2 dex.
Therefore, our SFR correlation would be even tighter if we had used time-averaged SFR.
In the bottom left corner are the intrinsically passive galaxies where the MIPS 24 $\mu m $ emission is coming solely from the evolved stars.
Most of them are predicted to have SFR not greater than $0.0001 \mathrm{M_\odot yr^{-1}}$.

\section{Broadband flux information}
\label{sec:broadb}
Table \ref{tab:bands} displays the selection of publicly available bands derived for each TNG50 galaxy with $\log M_{\mathrm{star}}>8 \; \mathrm{M_\odot}$.
\begin{table}
    \centering
    \caption{Available broadband fluxes. Survey, telescope or photometric system, band and its pivot wavelength.}
    \begin{tabular}{llc}
    \textbf{Survey/System} & \textbf{Band} & \textbf{Pivot wavelength [$\mu m$]}\\
    \hline
     \multirow{2}{*}{GALEX} &  FUV & 0.15\\
        & NUV & 0.23\\\hline
       \multirow{7}{*}{Johnson} & U & 0.35\\
         & B & 0.44\\
         & V & 0.55\\
         & R & 0.69\\
         & I & 0.87\\
         & J & 1.24\\
         & M & 5.01\\\hline
         \multirow{5}{*}{SDSS}  & u & 0.36\\
           & g & 0.47\\
           & r & 0.62\\
           & i & 0.75\\
           & z & 0.89\\\hline
       \multirow{5}{*}{UKIDSS} & Z & 0.88\\
        & Y & 1.03\\
        & J & 1.25\\
        & H & 1.64\\
        & K & 2.21\\\hline
         \multirow{3}{*}{2MASS}  & J & 1.24\\
        & H & 1.65\\
       & Ks & 2.16\\\hline
          \multirow{4}{*}{WISE} &  1 & 3.39\\
           & 2 & 4.64\\
           & 3 & 12.6\\
           & 4 & 22.3\\\hline
           \multirow{7}{*}{Spitzer} &   IRAC 1 & 3.55\\
          & IRAC 2 & 4.50\\
          & IRAC 3 & 5.72\\
          & IRAC 4 & 7.88\\
         & MIPS 24 & 23.8\\
         & MIPS 70 & 72\\
         & MIPS 160 & 156\\\hline
         \multirow{4}{*}{IRAS} &  12 & 11.4\\
           & 25 & 23.6\\
          & 60 & 60.4\\
          & 100 & 101\\\hline
        \multirow{6}{*}{Herschel} & PACS 70 & 70.8\\
         & PACS 100 & 101\\
         & PACS 160 & 162\\
        & SPIRE 250 & 253\\
        & SPIRE 350 & 354\\
        & SPIRE 500 & 515\\\hline
       \multirow{2}{*}{JCMT} & SCUBA2 450 & 449\\
       & SCUBA2 850  & 	854\\\hline
       \multirow{3}{*}{Planck} & 857 & 352\\
       & 545 & 545\\
       & 353 & 839\\\hline
       \multirow{5}{*}{ALMA} & 10 &  350\\
       & 9 & 456\\
        & 8 & 690\\
        & 7 & 938\\
        & 6 & 1244\\\hline
    \end{tabular}
    
    \label{tab:bands}
\end{table}

\section{Aperture definition}
\label{ape:petro}

\begin{figure}
	\includegraphics[width=0.9\columnwidth]{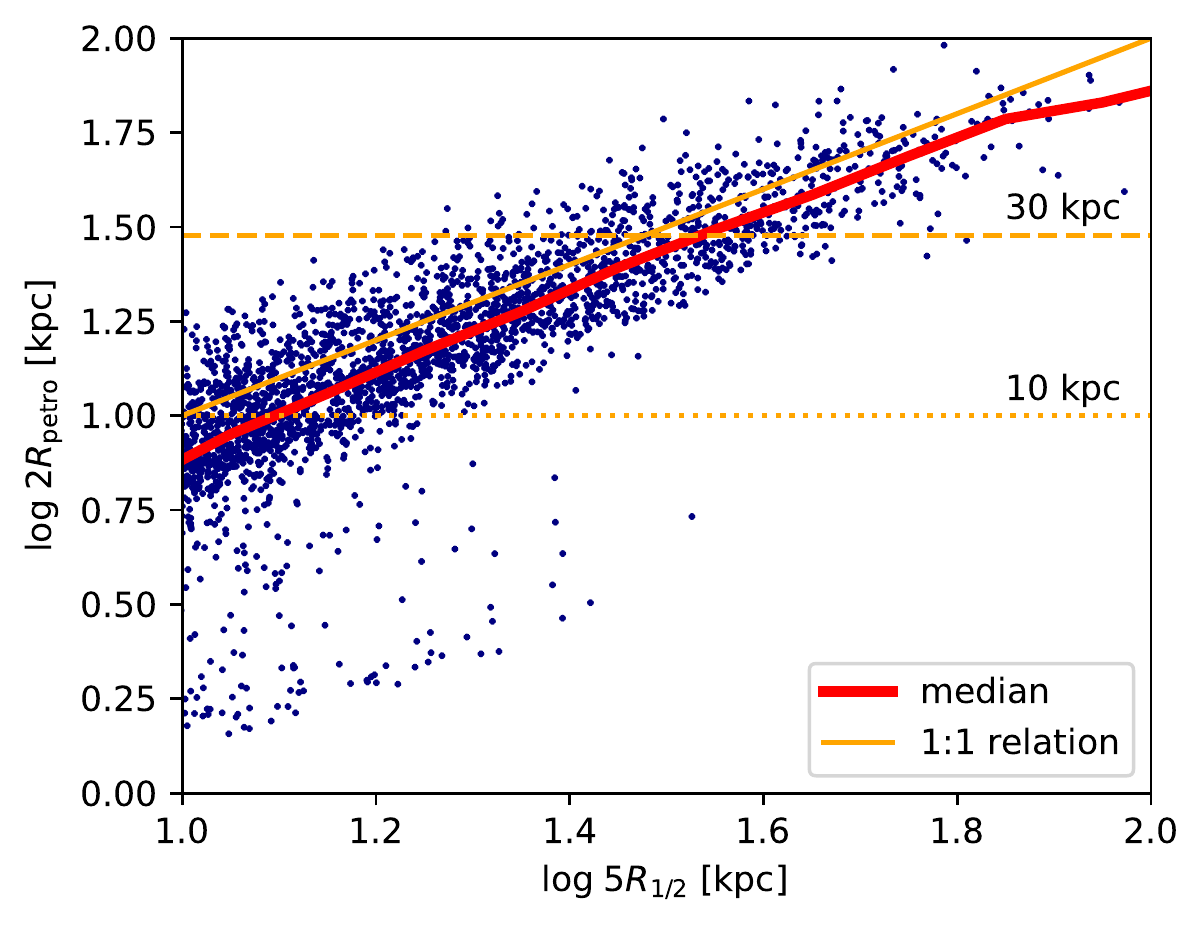}
     \caption{
     Correlation between our fiducial aperture (x-axis) and a Petrosian aperture, for a sub-sample of large TNG50 galaxies. The horizontal lines correspond to the 10 and 30 kpc apertures;  the orange solid line is one-to-one relation and the red line is the running median.}
    \label{fig:petro}
\end{figure}

To associate our aperture definitions with observational ones, we applied the \textsc{statmorph}\footnote{\href{https://statmorph.readthedocs.io/}{https://statmorph.readthedocs.io/}} package \citep{Rodriguez-Gomez2019} on images in the SDSS $i$ band, and derived the Petrosian radii ($R_{\mathrm{petro}}$) for all the galaxies in our sample. In Fig.~{\ref{fig:petro}} we show how our fiducial aperture at $z=0$ correlates with $2\,R_{\mathrm{petro}}$, the aperture typically used in the literature. Focusing on the range above 10 kpc, we see how the Petrosian aperture is slightly smaller than the fiducial one (no more than 0.1 dex), but the apertures are tightly correlated. While it seems that the aperture of 10 kpc mostly resolves the discrepancy between the observed and simulated LFs at the bright end, we see from Fig.~{\ref{fig:petro}} that for large galaxies (which populate the bright end of the LFs), this aperture will probably capture less light.
Instead, the aperture of 30 kpc is a better approximation, as also seen in \cite{Schaye2015, Pillepich2018b}, although they applied it in 3D while we did in 2D (along the line of sight, the height of the cylinder is always $10\,R_{\mathrm{1/2}}$).
They also focused on the larger simulation boxes, which include a higher number of the most massive/luminous galaxies.

\bsp	
\label{lastpage}
\end{document}